\documentclass{elsart}
\usepackage{graphicx,amssymb,natbib}
\def\aa{{\em A\&A}\ }

\def\aj{{\em AJ}\ }
\def\annrev{{\em ARA\&A}\ }
\def\apj{{\em ApJ}\ }
\def\apjs{{\em ApJS}\ }

\def\mnras{{\em MNRAS}\ }
\def\nat{{\em Nature}\ }
\def\nphys{{\em Nuclear Phys}\ }

\def\lsim{\mathrel{\rlap{\lower 4pt \hbox{\hskip 1pt $\sim$}}\raise 1pt
\hbox {$<$}}} 
\def\gsim{\mathrel{\rlap{\lower 4pt \hbox{\hskip 1pt $\sim$}}\raise 1pt
\hbox {$>$}}}
\newcommand{\ms}{$M_\odot$}
\newcommand{\Msun}{M_{\odot}}

\begin{document}
\runauthor{Nomoto, Tominaga, Umeda, Kobayashi, and Maeda}
\begin{frontmatter}

\title{Nucleosynthesis Yields of Core-Collapse Supernovae and
Hypernovae, and \\
Galactic Chemical Evolution}

\author[tokyo]{Ken'ichi Nomoto}
\author[tokyo]{Nozomu Tominaga}
\author[tokyo]{Hideyuki Umeda}
\author[nao]{Chiaki Kobayashi}
\author[tokyo2]{Keiichi Maeda}

\address[tokyo]{Department of Astronomy, University of Tokyo,
  Bunkyo-ku, Tokyo 113-0033, Japan}
\address[nao]{National Astronomical Observatory, Mitaka, Tokyo, Japan}
\address[tokyo2]{Department of Earth Science and Astronomy,
College of Arts and Science, University of Tokyo, Meguro-ku, Tokyo
153-8902, Japan}

\vspace{0.5cm}
{\normalsize {\it To appear in Nuclear Physics A (Special Issue on Nuclear
 Astrophysics) \\ eds. K. Langanke, F.-K. Thielemann, \& M. Wiescher
 (2006)}}

\begin{abstract}

We present new nucleosynthesis yields as functions of the stellar
mass, metallicity, and explosion energy (corresponding to normal
supernovae and Hypernovae).  We apply the results to the chemical
evolution of the solar neighborhood.  Our new yields are based on the
new developments in the observational/theoretical studies of
supernovae (SNe) and extremely metal-poor (EMP) stars in the halo,
which have provided excellent opportunities to test the explosion
models and their nucleosynthesis.  We use the light curve and spectra
fitting of individual SN to estimate the mass of the progenitor,
explosion energy, and produced $^{56}$Ni mass.  Comparison with the
abundance patterns of EMP stars has made it possible to determine the
model parameters of core-collapse SNe, such as mixing-fallback
parameters.

More specifically, we take into account the two distinct new classes
of massive SNe: 1) very energetic Hypernovae, whose kinetic energy
(KE) is more than 10 times the KE of normal core-collapse SNe, and 2)
very faint and low energy SNe (Faint SNe).  These two new classes of
SNe are likely to be ``black-hole-forming'' SNe with rotating or
non-rotating black holes.  Nucleosynthesis in Hypernovae is
characterized by larger abundance ratios (Zn,Co,V,Ti)/Fe and smaller
(Mn,Cr)/Fe than normal SNe, which can explain the observed trends of
these ratios in EMP stars.  Nucleosynthesis in Faint SNe is
characterized by a large amount of fall-back, which explains the
abundance pattern of the most Fe-poor stars.  These comparisons
suggest that black-hole-forming SNe made important contributions to
the early Galactic (and cosmic) chemical evolution.

\end{abstract}
\begin{keyword}
abundances; nucleosynthesis; Population III stars; supernovae
\end{keyword}
\end{frontmatter}

\section{Introduction}

Massive stars in the range of 8 to $\sim$ 130$M_\odot$ undergo
core-collapse at the end of their evolution and become Type II and
Ib/c supernovae (SNe) unless the entire star collapses into a black
hole with no mass ejection \cite[e.g.,][]{arnett1996, hillebrandt2003,
fryer2004}.  Here, supernovae are classified based on the maximum
light spectra as follows \cite[e.g.,][]{alex1997}.  Type II supernovae
(SNe II) are defined by the presence of hydrogen, which implies that
the progenitors are red (or blue) supergiants keeping their
hydrogen-rich envelope.  Type Ib supernovae (SNe Ib) are characterized
by the lack of hydrogen but the presence of prominent He lines, so
that their progenitors are Wolf-Rayet (WN) or He stars losing their
H-rich envelope in a stellar wind or by Roche lobe overflow in binary
systems.  Type Ic supernovae (SNe Ic) do not show prominent He lines
as well as H, which implies that their progenitors have lost even most
of He layers to become WC/WO Wolf-Rayet stars or C+O stars in binary
systems \cite[e.g.,][]{nomoto1994c}.  (In contrast, Type Ia SNe are
the thermonuclear explosions of mass accreting white dwarfs in binary
systems and spectroscopically characterized by the lack of H and He
and the presence of strong Si lines \cite[e.g.,][]{nomoto1994a}.)

These supernovae release large explosion energies and eject explosive
nucleosynthesis materials, thus having strong dynamical, thermal, and
chemical influences on the evolution of interstellar, intergalactic,
and intracluster matter \cite[e.g.,][]{matteucci2002} as well as
galaxies and galaxy clusters.  Therefore, the explosion energies of
core-collapse supernovae are fundamentally important quantities, and
an estimate of $E \sim 1\times 10^{51}$ ergs has often been used in
calculating nucleosynthesis and the impact on the interstellar medium.
(In the present paper, we use the explosion energy $E$ for the final
kinetic energy of explosion, and $E_{51} = E/10^{51}$\,erg.)  A good
example is SN1987A in the Large Magellanic Cloud, whose energy is
estimated to be $E_{51} = 1.0 - 1.5$ from its early light curve
\cite[e.g.,][]{arnett1996, nomoto1994b}.

One of the most interesting recent developments in the study of
supernovae is the discovery of some very energetic supernovae, whose
kinetic energy (KE) exceeds $10^{52}$\,erg, more than 10 times the KE
of normal core-collapse SNe.  The most luminous and powerful of these
objects, the Type Ic supernova (SN~Ic) 1998bw, was linked to the
gamma-ray burst GRB 980425 \cite{galama1998}, thus establishing for
the first time a connection between gamma-ray bursts (GRBs) and the
well-studied phenomenon of core-collapse SNe \cite{woosley1993,
pac1998, woosley2006}.  However, SN~1998bw was exceptional for a
SN~Ic: it was as luminous at peak as a SN~Ia, indicating that it
synthesized $\sim 0.5 M_\odot$ of $^{56}$Ni, and its KE was estimated
at $E_{51} \sim 30$ \cite{iwamoto1998}.

In the present paper, we use the term 'Hypernova (HN)' to describe
such a hyper-energetic supernova with $E \gsim 10^{52}$ ergs without
specifying the explosion mechanism \cite{nomoto2001}.  Following SN
1998bw, other ``hypernovae'' of Type Ic have been discovered or
recognized \cite{nomoto2004}.  

Nucleosynthesis features in such hyper-energetic (and
hyper-aspherical) supernovae must show some important differences from
normal supernova explosions.  This might be related to the unpredicted
abundance patterns observed in the extremely metal-poor (EMP) halo
stars \cite[e.g.,][]{hill2005, beers2005}.  This approach leads to
identifying the First Stars in the Universe, i.e., metal-free,
Population III (Pop III) stars which were born in a primordial
hydrogen-helium gas cloud.  This is one of the important challenges of
the current astronomy \cite[e.g.,][]{weiss2000, abel2002, bromm2004}.

More generally, the enrichment by a single SN can dominate the
preexisting metal contents in the early universe \cite[e.g.,][]{aud95,
ryan1996, shigeyama1998, nakamura1999}.  Therefore, the comparison
between the SN model and the abundance patterns of EMP stars can
provide a new way to find out the individual SN nucleosynthesis.

In \S2, we briefly describe how the progenitor mass $M$ and explosion
energy $E$ are estimated from the observations of
supernovae/hypernovae.  In \S3, the characteristics of nucleosynthesis
in hypernovae are investigated with detailed nucleosynthesis
calculations and compared with nucleosynthesis in normal supernovae.
In \S4, we then discuss possible contribution of hypernovae to the
Galactic chemical evolution and on the abundances in metal-poor stars.

\section{Hypernovae and Faint Supernovae}

\begin{figure*}
\centering
\includegraphics*[width=9.5cm]{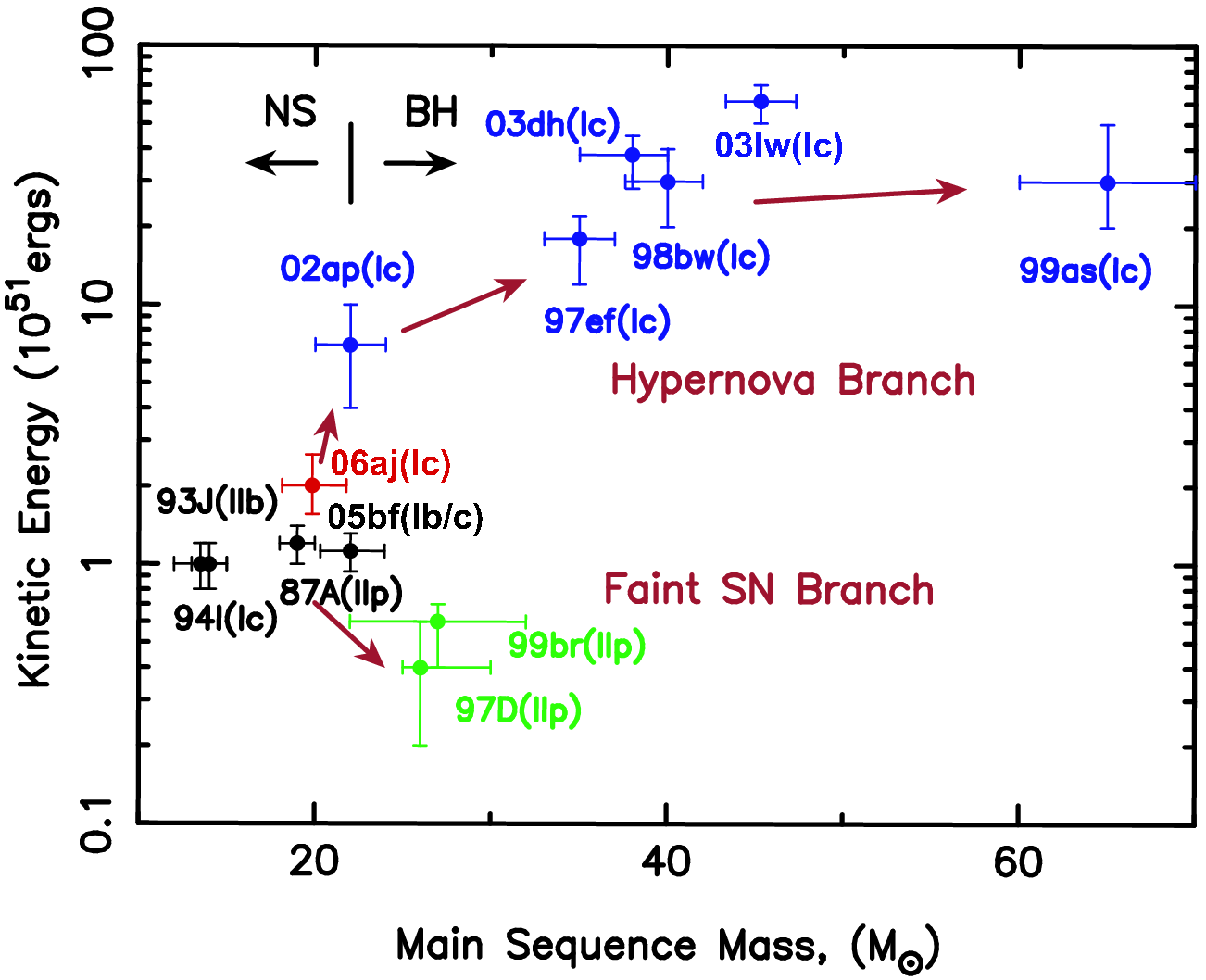}
\includegraphics*[width=9.5cm]{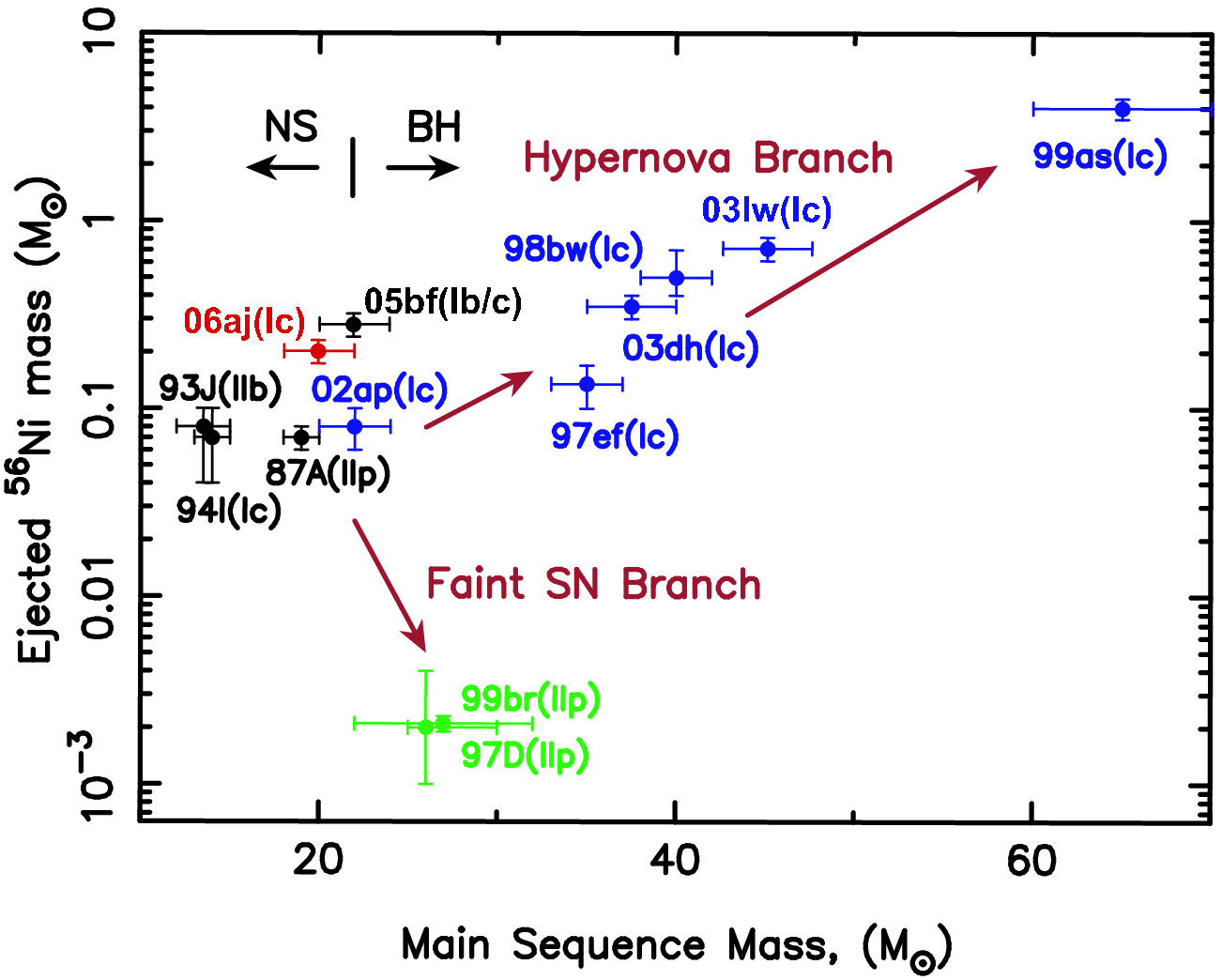}
\caption{
The explosion energy and the ejected $^{56}$Ni mass as a function of
the main sequence mass of the progenitors for several
supernovae/hypernovae.}
\label{fig2}
\end{figure*}

The connection between long GRBs and core-collapse SNe has been
clearly established from GRB 980425/SN 1998bw \cite{galama1998}, GRB
030329/SN 2003dh \cite{stanek2003, hjorth2003}, and GRB 031203/SN
2003lw \cite{mal04}.  As summarized in Figure~\ref{fig2}, these
GRB-SNe have similar properties; they are all hypernovae with $E_{51}
\sim$ 30 - 50 and synthesize 0.3 - 0.5 $M_\odot$ of $^{56}$Ni
\cite{iwamoto1998, nakamura2001a, mazzali2003}.

Hypernovae are also characterized by asphericity from the observations
of polarization and emission line features \cite[e.g.,][]{wang2003,
kawabata2002, maeda2002}.  The explosion energy of the aspherical
models for hypernovae tends to be smaller than the spherical models by
a factor of 2 - 3, but still being as high as $E_{51} \gsim 10$
\cite{mae03}.

Recently X-Ray Flash (XRF) 060218 has been found to be connected to SN
Ic 2006aj \cite{campana2006, pian2006}.  Compared with the above
GRB-SNe, SN 2006aj is a less energetic ($E_{51} \sim 2$) SN from a
smaller mass progenitor, $\sim 20 M_\odot$, thus being suggested to be
a ``neutron star-making SN'' \cite{mazzali2006}.

Other non-GRB ``hypernovae'' have been recognized, such as SN~1997ef
\cite{iwamoto2000, mazzali2000} and SN~2002ap \cite{mazzali2002}.
These hypernovae span a wide range of properties, although they all
appear to be highly energetic compared to normal core-collapse SNe.
The mass estimates, obtained from fitting the optical light curves and
spectra, place hypernovae at the high-mass end of SN progenitors.

In contrast, SNe II 1997D and 1999br were very faint SNe with very low
KE \cite{turatto1998, hamuy2003, zampieri2003}.  In the diagram that
shows $E$ and the mass of $^{56}$Ni ejected $M(^{56}$Ni) as a function
of the main-sequence mass $M_{\rm ms}$ of the progenitor star
(Figure~\ref{fig2}), therefore, we propose that SNe from stars with
$M_{\rm ms} \gsim 20-25 M_\odot$ have different $E$ and $M(^{56}$Ni),
with a bright, energetic ``hypernova branch'' at one extreme and a
faint, low-energy SN branch at the other \cite{nomoto2003}.  For the
faint SNe, the explosion energy was so small that most $^{56}$Ni fell
back onto the compact remnant \cite[e.g.,][]{sollerman1998}.  Thus the
faint SN branch may become a ``failed'' SN branch at larger $M_{\rm
ms}$.  Between the two branches, there may be a variety of SNe
\cite{hamuy2003, tominaga2005}.

This trend might be interpreted as follows.  Stars more massive than
$\sim$ 25 $M_\odot$ form a black hole at the end of their evolution.
Stars with non-rotating black holes are likely to collapse ``quietly''
ejecting a small amount of heavy elements (Faint supernovae).  In
contrast, stars with rotating black holes are likely to give rise to
Hypernovae.  The hypernova progenitors might form the rapidly rotating
cores by spiraling-in of a companion star in a binary system.

\section{Nucleosynthesis in Hypernovae}

Here we summarize nucleosynthesis yields of core-collapse SN models
for 13 -- 40 $M_\odot$ stars for various explosion energies and
progenitor metallicity.  Compared with earlier works \cite{woo95,
tnh96, nomoto1997, rauscher2002, cl2002, lc2003}, we focus on (1)
Hypernova models to compare with the GRB-SNe and other observations,
and (2) Pop III SN models in order to compare with observed abundance
patterns of EMP stars.

For HNe, we adopt the ($M_{\rm ms}-E$) relation as estimated from
observations and models of SNe (Fig.~\ref{fig2}), i.e., for $M_{\rm
ms} = 20, 25, 30, 40, 50 M_\odot$, $E_{51}=$ 10, 10, 20, 30, 40,
respectively.  For normal SNe II of 13 - 50 $M_\odot$, $E_{51}=1$ is
assumed \cite{tominaga2006}.

The new ingredients taken into account in the present nucleosynthesis
models are:
\noindent
(i) the variation of $E$ (hypernovae, normal SNe, and faint SNe),
\noindent
(ii) the {\sl mixing and fallback}, and
\noindent
(iii) neutrino processes that affects neutron excess near the mass
cut.

\subsection {Energy Dependence}

In core-collapse supernovae/hypernovae, stellar material undergoes
shock heating and subsequent explosive nucleosynthesis. Iron-peak
elements are produced in two distinct regions, which are characterized
by the peak temperature, $T_{\rm peak}$, of the shocked material.  For
$T_{\rm peak} > 5\times 10^9$K, material undergoes complete Si burning
whose products include Co, Zn, V, and some Cr after radioactive
decays.  For $4\times 10^9$K $<T_{\rm peak} < 5\times 10^9$K,
incomplete Si burning takes place and its after decay products include
Cr and Mn \cite[e.g.,][]{nakamura1999}.

\begin{figure*}
\centering
\includegraphics*[width=6.5cm]{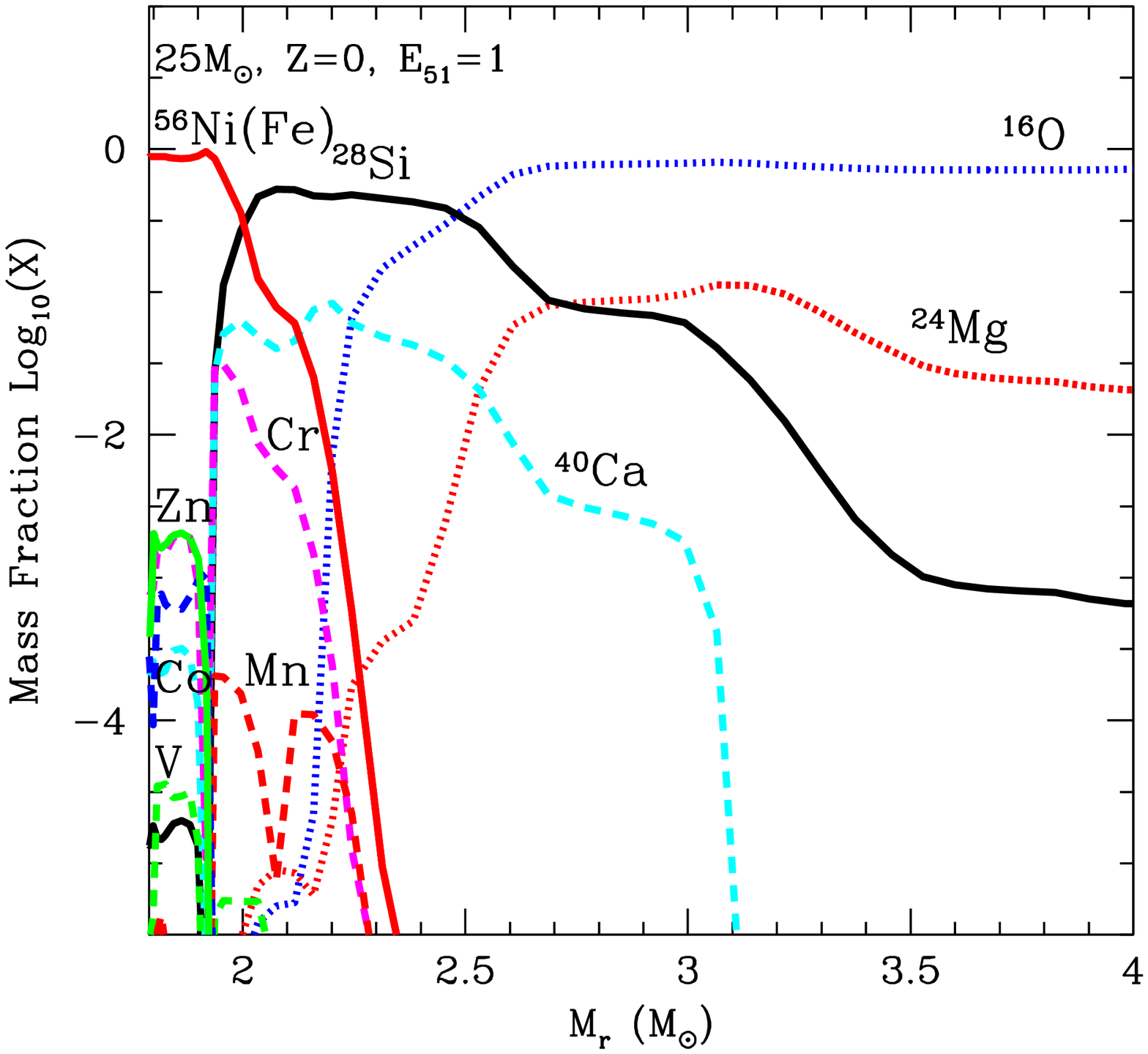}
\includegraphics*[width=6.5cm]{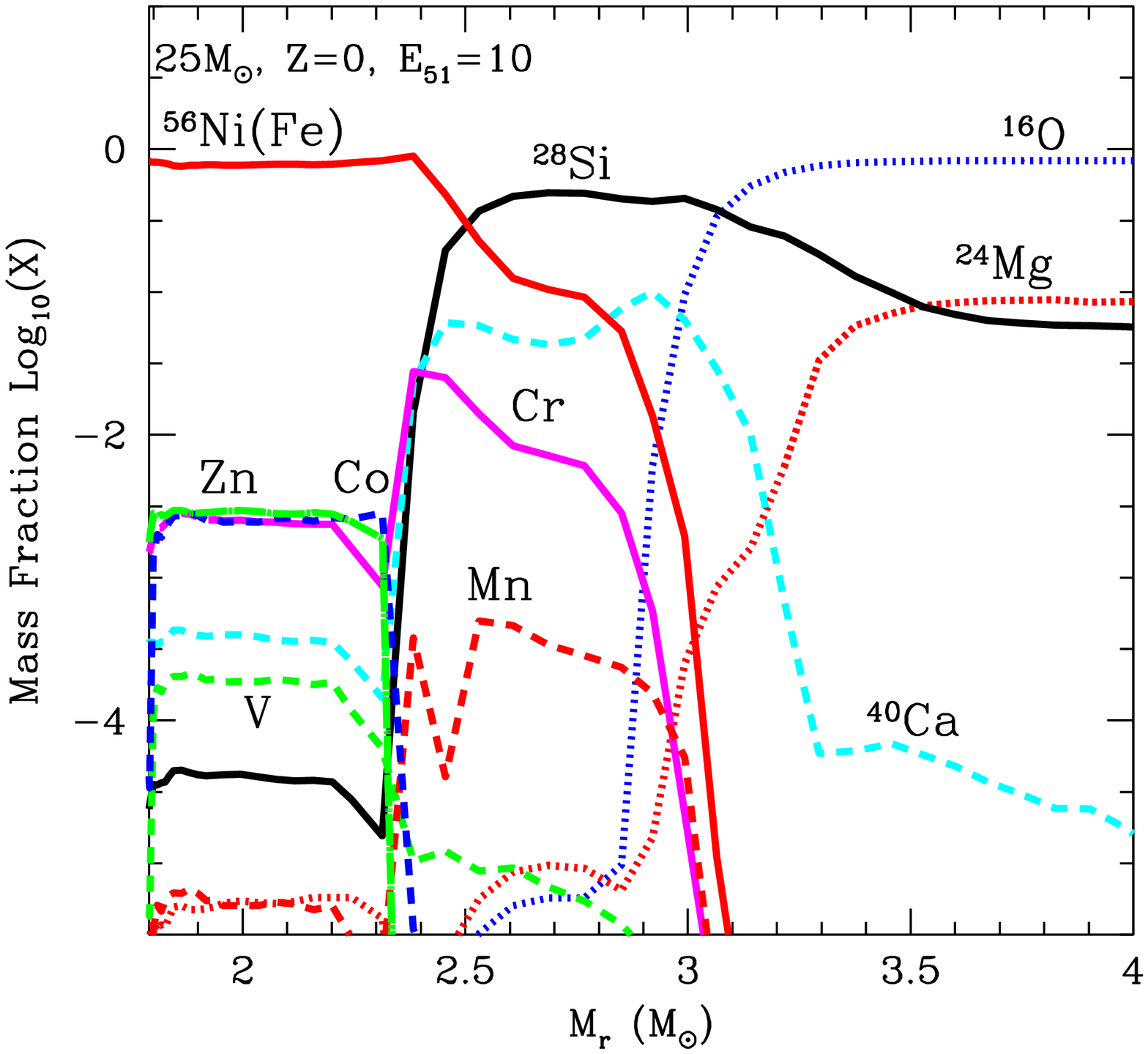}
\caption{Abundance distribution against the enclosed mass
$M_r$ after the explosion of Pop III 25 \ms\ stars with $E_{51} = 1$
(top) and $E_{51} = 10$ (bottom) \cite{umeda2002a}.}
\label{fig3}
\end{figure*}	

The right panel of Figure~\ref{fig3} shows the composition in the
ejecta of a 25 $M_\odot$ hypernova model ($E_{51} = 10$).  The
nucleosynthesis in a normal 25 $M_\odot$ SN model ($E_{51} = 1$) is
also shown for comparison in the left panel of Figure~\ref{fig3}
\cite{umeda2002a}.

We note the following characteristics of nucleosynthesis with very
large explosion energies \cite{nakamura2001b, nomoto2001, umeda2005}:

(i) Both complete and incomplete Si-burning regions shift outward in
mass compared with normal supernovae, so that the mass ratio between
the complete and incomplete Si-burning regions becomes larger.  As a
result, higher energy explosions tend to produce larger [(Zn, Co,
V)/Fe] and smaller [(Mn, Cr)/Fe], which can explain the trend observed
in very metal-poor stars \cite{umeda2005}.
(Here [A/B] $= \log_{10}(N_{\rm A}/N_{\rm B})-\log_{10} (N_{\rm
A}/N_{\rm B})_\odot$, where the subscript $\odot$ refers to the solar
value and $N_{\rm A}$ and $N_{\rm B}$ are the abundances of elements A
and B, respectively.)

(ii) In the complete Si-burning region of hypernovae, elements produced
by $\alpha$-rich freezeout are enhanced.  Hence, elements synthesized
through capturing of $\alpha$-particles, such as $^{44}$Ti, $^{48}$Cr,
and $^{64}$Ge (decaying into $^{44}$Ca, $^{48}$Ti, and $^{64}$Zn,
respectively) are more abundant.

(iii) Oxygen burning takes place in more extended regions for the larger
KE.  Then more O, C, Al are burned to produce a larger amount of
burning products such as Si, S, and Ar.  Therefore, hypernova
nucleosynthesis is characterized by large abundance ratios of
[Si,S/O], which can explain the abundance feature of M82 
\cite{umeda2002b}.

\begin{figure*}
\includegraphics*[width=6.5cm]{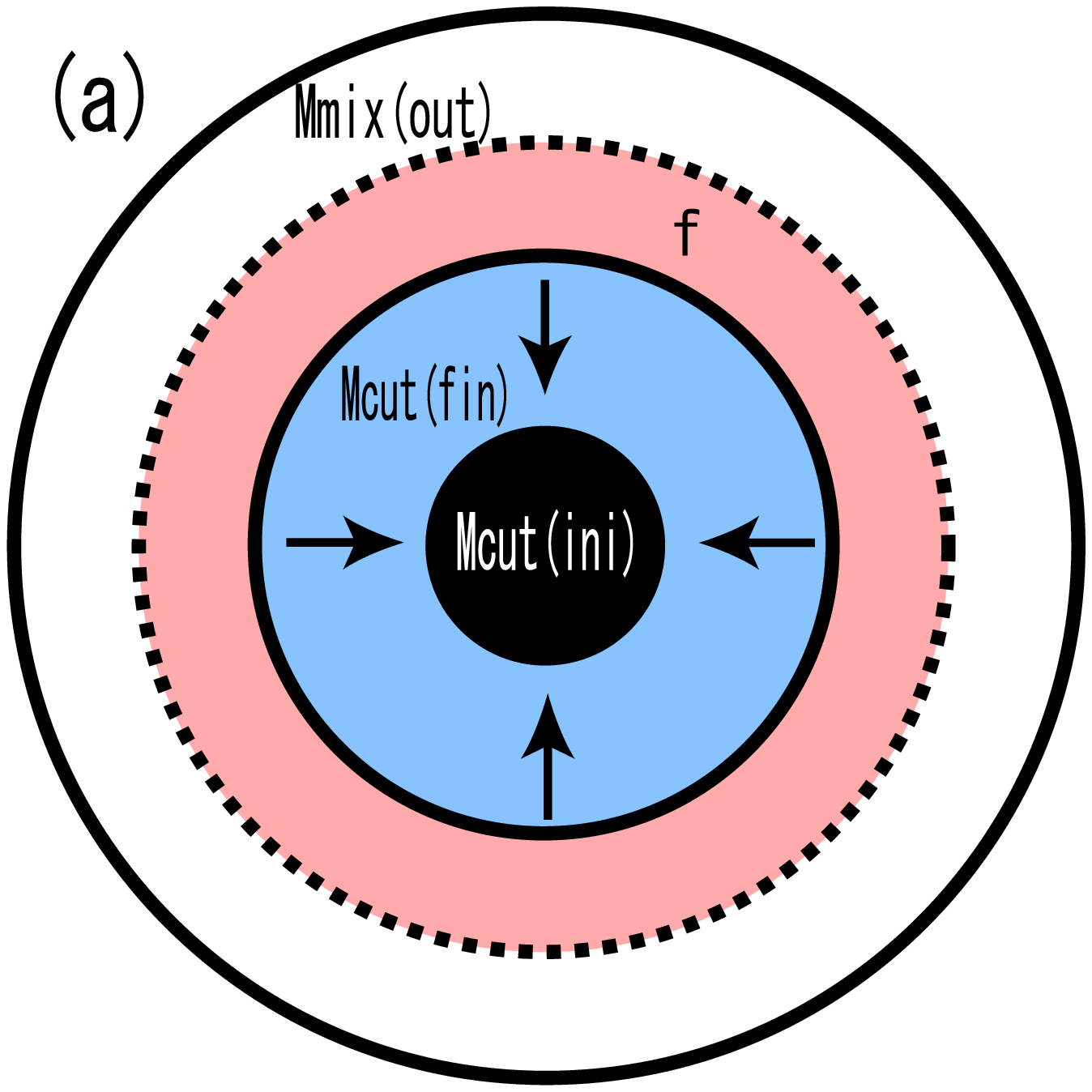}
\includegraphics*[width=6.5cm]{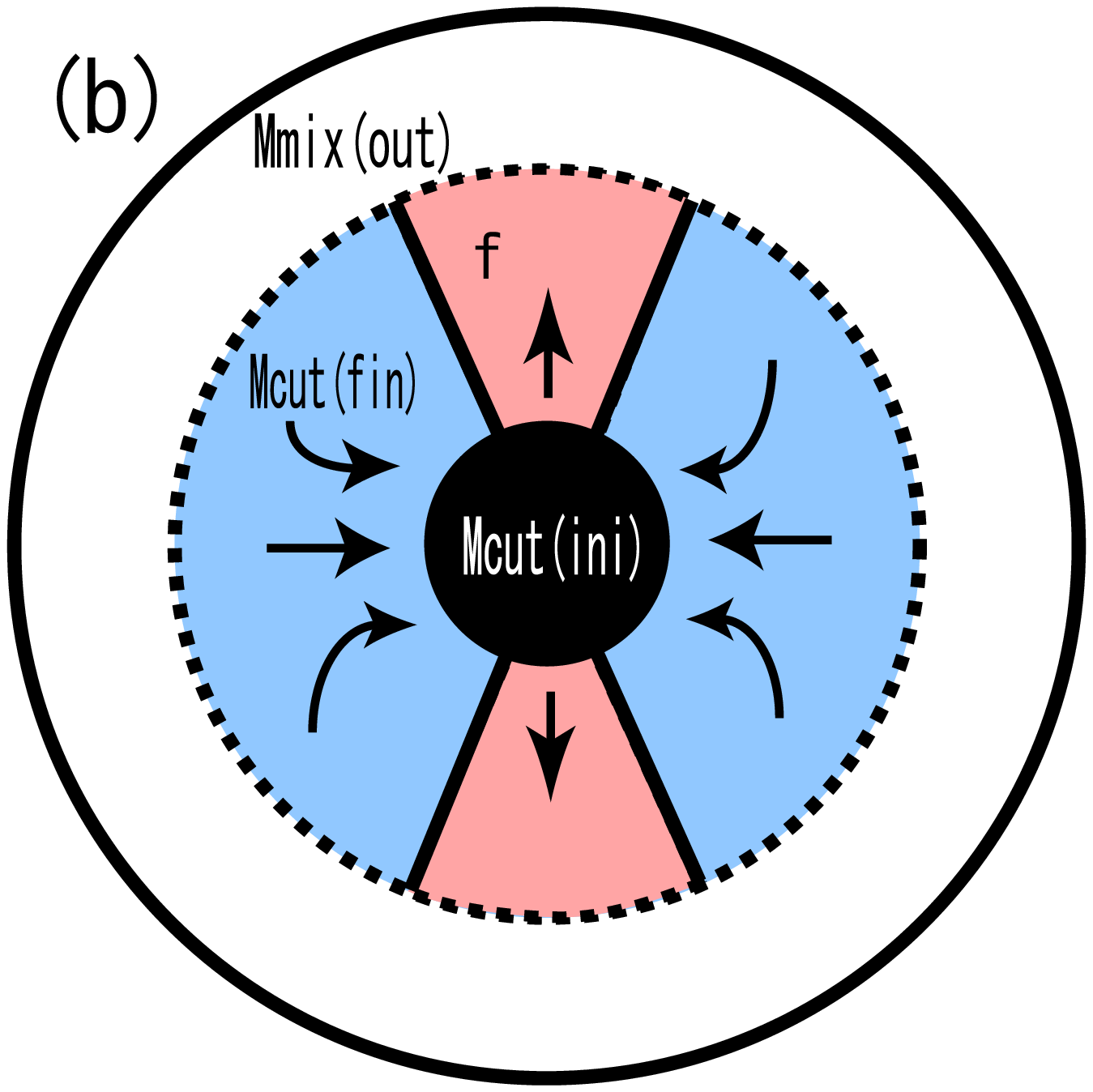}

\caption{The illustration of the mixing-fallback model.  The central
black region is the initial mass cut, that is, inside the inner
boundary of the mixing region, $M_{\rm cut}$(ini).  The mixing
region is enclosed with the dotted line at $M_{\rm mix}$(out).
A fraction $f$ of the materials in the mixing region is ejected into
the interstellar space.  The rest materials, locating in the gray
region inside of $M_{\rm cut}$(fin), undergo fallback onto the
central remnant.  (a) 1-dimensional picture: The materials mixed up to
a given radius, and a part of the materials are ejected.  (b)
2-dimensional picture: While all materials in the outer region above
$M_{\rm mix}{\rm (out)}$ are ejected, the materials in the mixing
region may be ejected only along the jet-axis.  In the jet-like
explosion, the ejection factor $f$ depends on the jet-parameters
(e.g., an opening angle and an energy injection rate).}
\label{fig:MF}
\end{figure*}

\subsection{Mixing-Fallback Model}

The mixing-fallback models are illustrated in Figures~\ref{fig:MF}ab.
First, the inner materials at $M_{\rm cut}{\rm (ini)} \le M_r \le
M_{\rm mix}{\rm (out)}$ are assumed to be mixed by Rayleigh-Taylor
instabilities and/or aspherical explosions during the shock wave
propagations in the star \cite[e.g.,][]{hac90, kif03}.  In fact, the
post-shock materials are convectively unstable because of
deceleration.  Later, some fraction of materials in the mixing region
undergoes fallback onto the central remnant by gravity
\cite[e.g.,][]{woo95, iwamoto2005}, and the rests at $M_r \ge M_{\rm
cut}{\rm (fin)}$ are ejected into interstellar space
(Fig.~\ref{fig5}).  The yields are made from the materials which exist
above the mixing region and which do not fall back. The degree of
fallback depends on the explosion energy, the gravitational potential,
and asphericity.

Fallback can take place not only for relatively low energy explosions
but also for very energetic jet-like explosions. In fact, Maeda et
al. \cite{mae03} have simulated jet-like explosions and showed that
their yields can be similar to the mixing-fallback model because the
materials around the jet axis are ejected and the materials around the
equatorial plane fallback onto the central remnant. The jet-like
explosion is consistent with a recently observed association between
HNe and GRBs with highly collimated relativistic jets. The
mixing-fallback model mimics such aspherical explosions, although the
spherical model tends to require larger explosion energies than the
jet model to obtain similar yields \cite{mae03}.

\subsection{Neutron-Proton Ratio near Mass Cut}

Recent studies \cite[e.g.,][]{rampp2000, janka2001, liebend03,
liebend05, frohlich06, burrows2006} have suggested that $Y_e$ may be
significantly varied by the neutrino process during explosion. The
region, where the neutrino absorption and the $Y_e$ variation occur,
is Rayleigh-Taylor unstable, thus having a large uncertainty in $Y_e$
distribution.

In most of present models, following $Y_e$ profile is adopted:
$Y_e=0.5001$ in the complete Si burning region and $Y_e=0.4997$ in the
incomplete Si burning region \cite{umeda2005}.  Lower $Y_e$ in the
incomplete Si-burning region leads to larger Mn/Fe, and larger $Y_e$
in the complete Si-burning region leads to even larger Co/Fe.

\begin{figure*}
\includegraphics*[width=12cm]{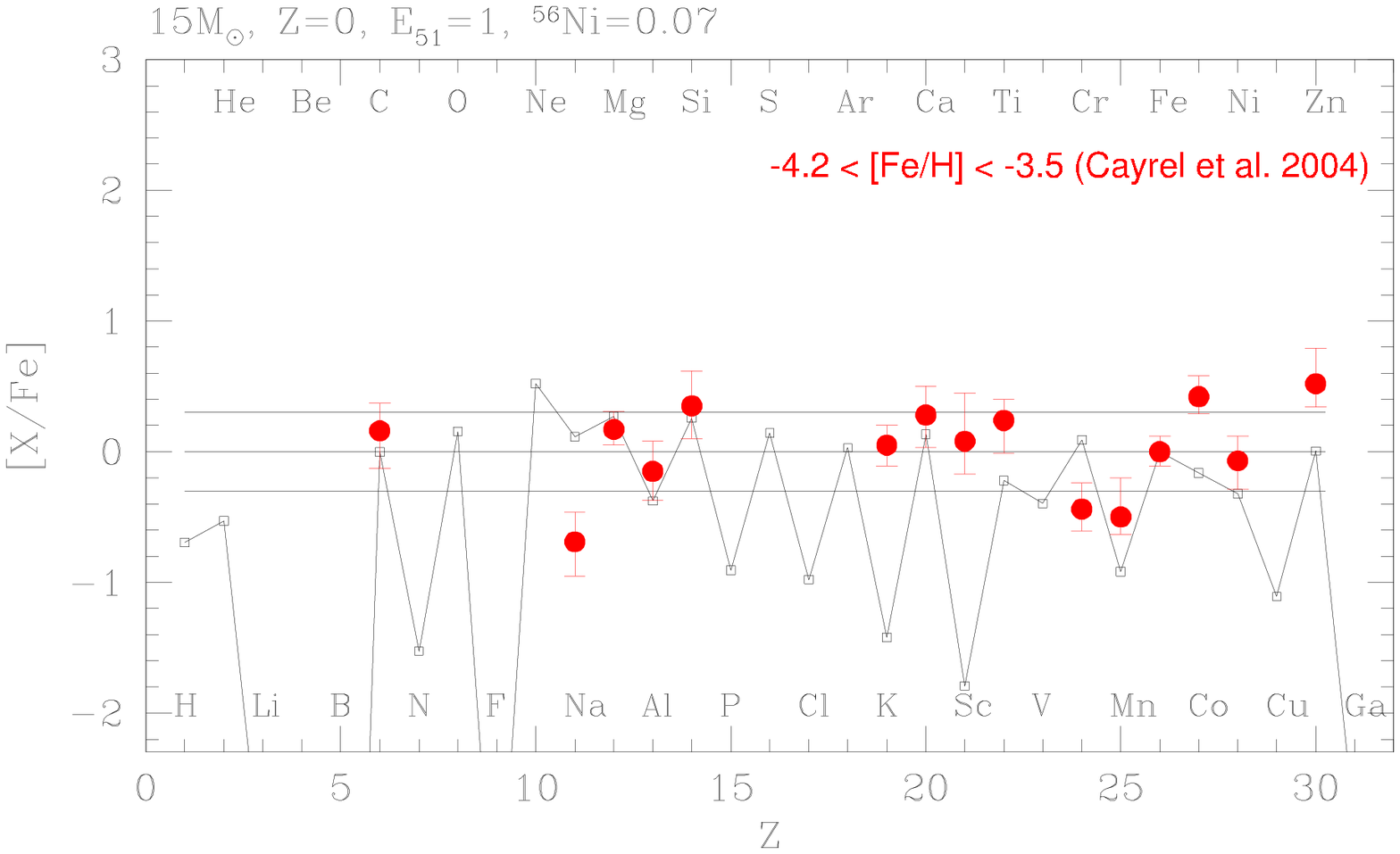}
\includegraphics*[width=12cm]{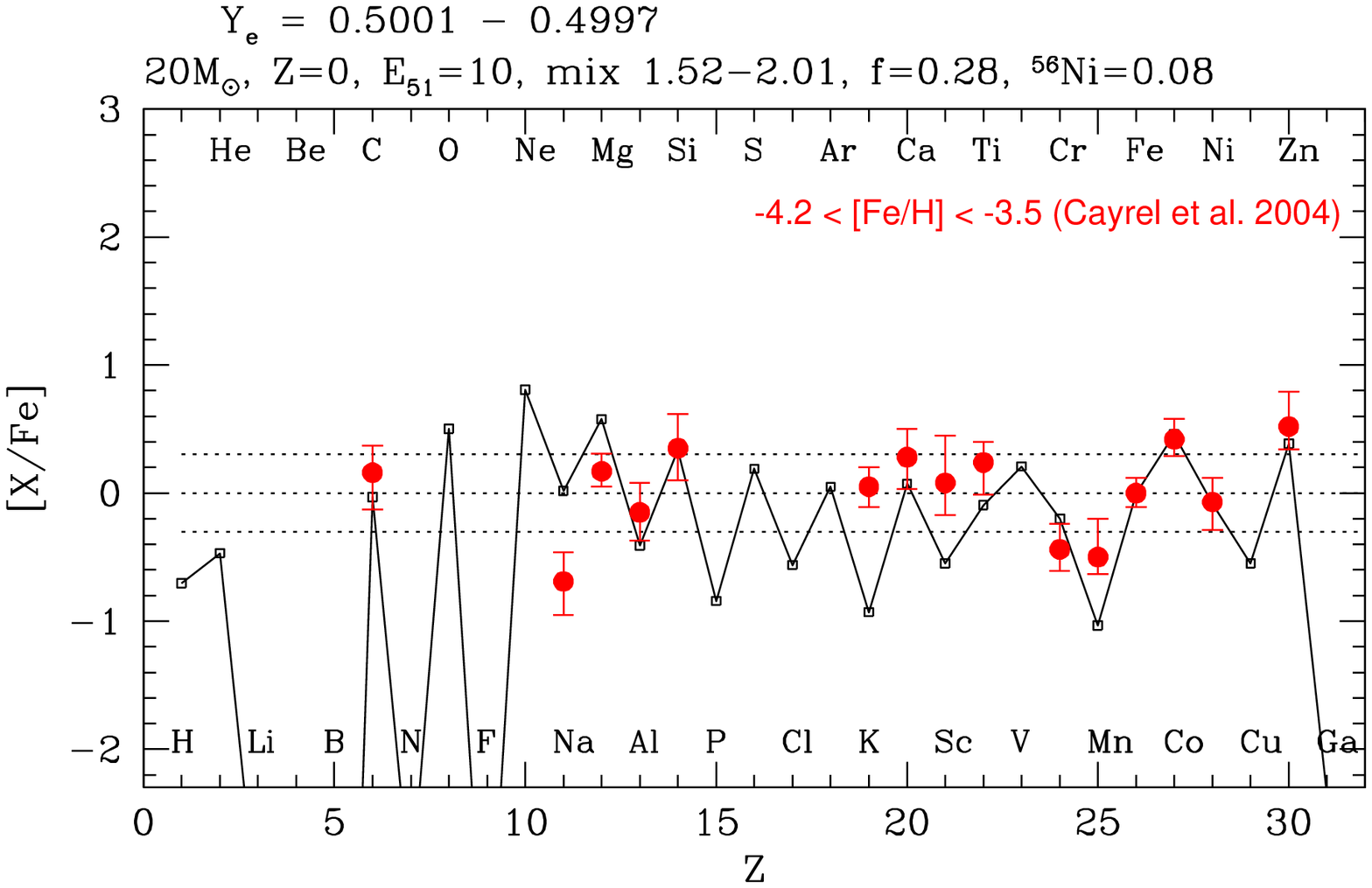}
\caption{Averaged elemental abundances of stars with [Fe/H] $= -3.7$
\cite{cayrel2004} compared with the normal SN yield (upper: 15
$M_\odot$, $E_{51} =$ 1) and the hypernova yield (lower: 20 $M_\odot$,
$E_{51} =$ 10).}
\label{fig7}
\end{figure*}

\section{Comparison with the Abundance Patterns of Metal-Poor Stars}

Cayrel et al. \cite{cayrel2004} provided abundance patterns of 35
metal-poor stars with small error bars for $-4.2\lsim{\rm
[Fe/H]}\lsim-2.0$.  Among the observed abundances, CNO elements might
be affected by the deep mixing in the evolved metal-poor stars
themselves, which itself is very interesting.  On the other hand, the
abundance patterns of heavier elements, e.g., Fe-peak elements, must
reflects the abundances of interstellar gases out of which the stars
were formed, thus providing us with the earliest chemical enrichment
in the galaxy.  These patterns have provided excellent materials to
compare with the SN/HN nucleosynthesis yields.

\subsection{Very Metal-Poor (VMP) Stars}

\begin{figure*}
\includegraphics*[width=12cm]{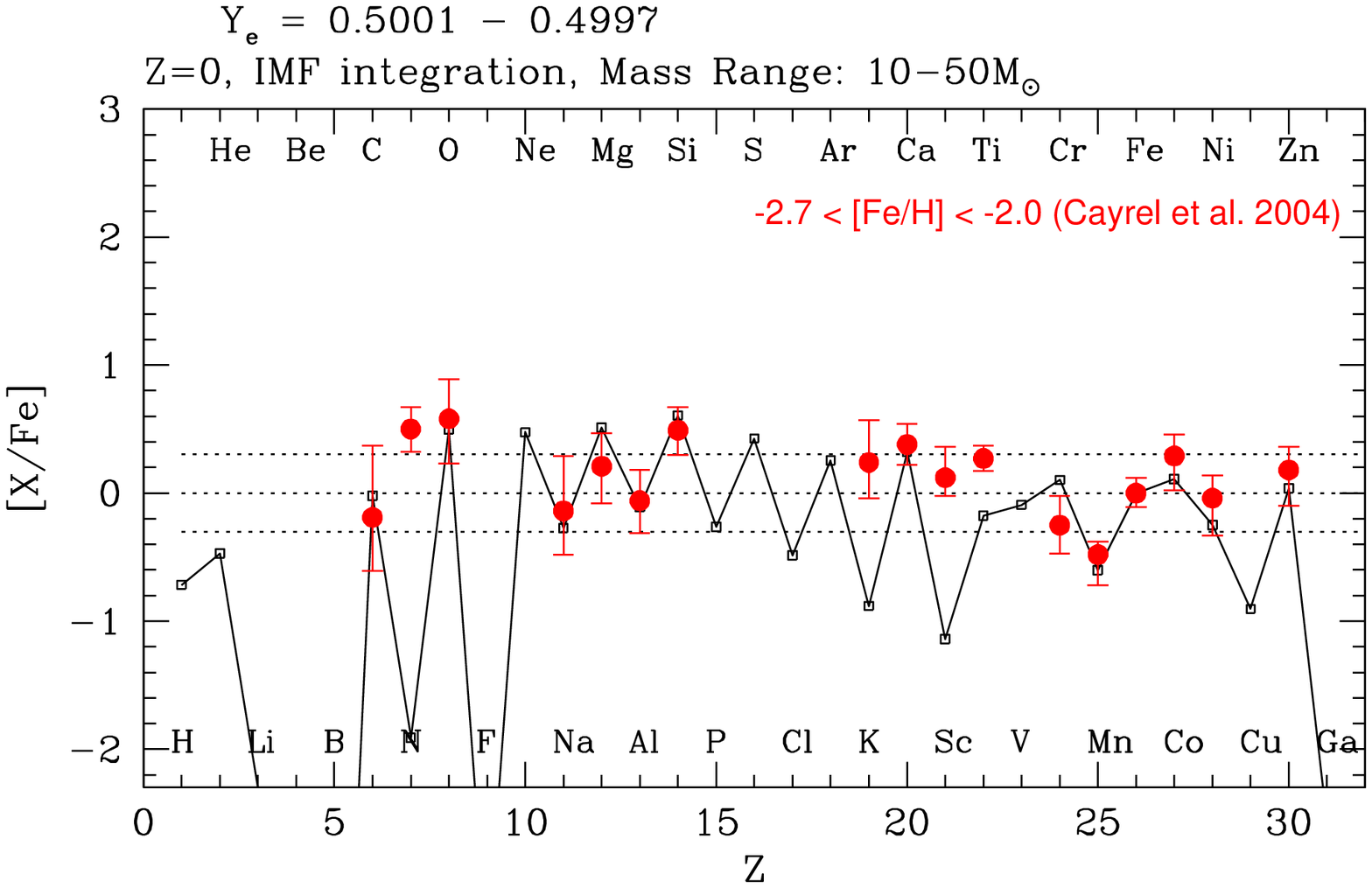}
\caption{Comparison between the abundance pattern of VMP stars
\cite{cayrel2004} ({\it filled circles with error bars}) and the IMF
integrated yield of Pop III SNe from 10$\Msun$ to 50 $\Msun$
\cite{tominaga2006}
}
\label{fig:IMF}
\end{figure*}

VMP stars defined as [Fe/H] $\lsim -2.5$ \cite{beers2005} are likely
to have the abundance pattern of well-mixed ejecta of many SNe.  We
thus compare the abundance patters of VMP stars with the SN yields
integrated over the progenitors of 10 - 50 $M_\odot$
(Fig.~\ref{fig:IMF}).

Since the abundance patterns of supernova models with [Fe/H] $\lsim
-2.5$ are quite similar to those of Pop III star models
\cite{umeda2000, woo95}, we use the Pop III yields for VMP and EMP
stars.  Comparison between the integrated yields over the Salpeter's
IMF and the abundance pattern of VMP stars (Fig.~\ref{fig:IMF}) show
that many elements are in reasonable agreements.

Figure~\ref{fig:IMF} shows that N is underproduced in these models.
There are two possible explanations for this discrepancy:

\noindent
(i) N was underproduced in the Pop III SN as in these models, but was
enhanced as observed during the first dredge-up in the low-mass
red-giant EMP stars \cite[e.g.,][]{weiss2004, suda2004}.  Actually,
most EMP stars are red-giants.

\noindent
(ii) N was enhanced in massive progenitor stars before the SN
explosion.  N is mainly synthesized by the mixing between the He
convective shell and the H-rich envelope \cite[e.g.,][]{umeda2000,
  iwamoto2005}.  Mixing can be enhanced by rotation
\cite{langer1992, heger2000, maeder2000}.  Suppose that the Pop III SN
progenitors were rotating faster than more massive stars because of
smaller mass loss, then [N/Fe] was enhanced as observed in EMP stars.

\subsection{Extremely Metal-Poor (EMP) Stars}

In the early galactic epoch when the galaxy was not yet chemically
well-mixed, each EMP star may be formed mainly from the ejecta of a
single Pop III SN (although some of them might be the second or later
generation SNe) \cite[e.g.,][]{argast2000, tum06}.  The formation of
EMP stars was driven by a supernova shock, so that [Fe/H] was
determined by the ejected Fe mass and the amount of circumstellar
hydrogen swept-up by the shock wave \cite{ryan1996}.  Then, hypernovae
with larger $E$ are likely to induce the formation of stars with
smaller [Fe/H], because the mass of interstellar hydrogen swept up by
a hypernova is roughly proportional to $E$ \cite{ryan1996,
shigeyama1998} and the ratio of the ejected iron mass to $E$ is
smaller for hypernovae than for normal supernovae.

The theoretical yields are compared with the averaged abundance
pattern of four EMP stars, CS~22189-009, CD-38:245, CS~22172-002 and
CS~22885-096, which have low metallicity ($-4.2<{\rm [Fe/H]}<-3.5$)
and normal [C/Fe] $\sim 0$ \cite{cayrel2004}.

Figure~\ref{fig7} shows that the averaged abundances of EMP stars can
be fitted well with the hypernova model of 20 $M_\odot$ and $E_{51} =$
10 (lower) but not with the normal SN model of 15 $M_\odot$ and
$E_{51} =$ 1 (upper) \cite{nomoto2005, tominaga2006}.

In the normal SN model (upper), the mass-cut is determined to eject Fe
of mass 0.14 $M_\odot$).  Then the yields are in reasonable agreements
with the observations for [(Na, Mg, Si)/Fe], but give too small [(Mn,
Co, Ni, Zn)/Fe] and too large [(Ca, Cr)/Fe].

In the HN model (lower), these ratios are in much better agreement
with observations.  The ratios of Co/Fe and Zn/Fe are larger in higher
energy explosions since both Co and Zn are synthesized in complete Si
burning at high temperature region (see the next subsection).  To
account for the observations, materials synthesized in a deeper
complete Si-burning region should be ejected, but the amount of Fe
should be small.  This is realized in the mixing-fallback models
\cite{umeda2002a, umeda2005}.

\begin{figure*}[!ht]
\centering
\includegraphics*[width=14cm]{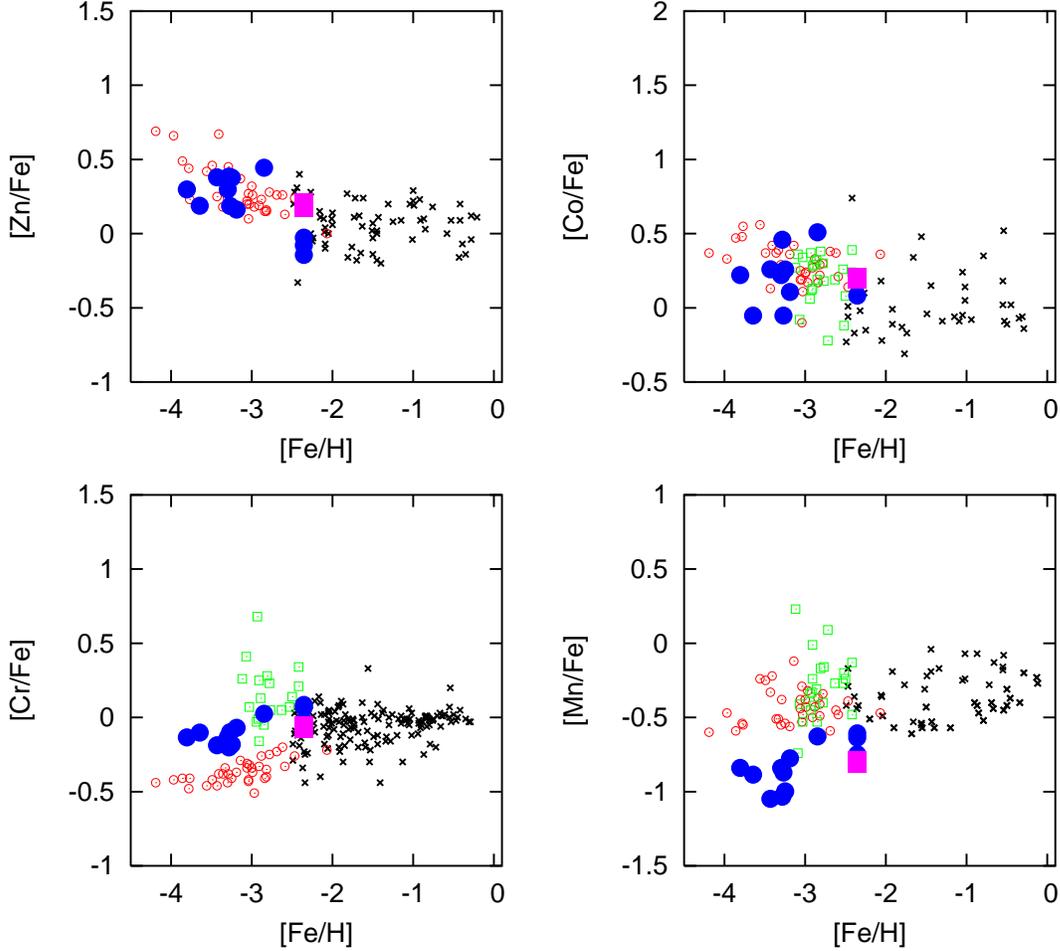}
\caption{Observed abundance ratios of [Zn, Co, Cr, Mn/Fe] vs [Fe/H]
[{\it open circle}: \cite{cayrel2004}; {\it open square}: \cite{hon04}]
compared with individual Pop III SN models ({\it filled circle}) and
 IMF-integrated models ({\it filled square}).}
\label{fig:trend}
\end{figure*}

\subsection{Hypernovae and Fe-peak Elements (Cr, Mn, Fe, Zn, Co, Zn)}

In the observed abundances of halo stars, there are significant
differences between the abundance patterns in the iron-peak elements
below and above [Fe/H]$ \sim -2.5$ - $-3$.

\noindent

(i) For [Fe/H]$\lsim -2.5$, the mean values of [Cr/Fe] and [Mn/Fe]
decrease toward smaller metallicity, while [Co/Fe] increases
\cite{mcw95, ryan1996}.  
[The negligibly weak trend in [Mn/Fe] in \cite{cayrel2004} is
different from previous observations that [Mn/Fe] decreases
significantly for smaller [Fe/H] \cite{mcw95}.]

\noindent
(ii) [Zn/Fe]$ \sim 0$ for [Fe/H] $\simeq -3$ to $0$ \cite{sneden1991},
while at [Fe/H] $< -3.3$, [Zn/Fe] increases toward smaller metallicity
\cite{cayrel2004}.

The larger [(Zn, Co)/Fe] and smaller [(Mn, Cr)/Fe] in the supernova
ejecta can be realized if the mass ratio between the complete Si
burning region and the incomplete Si burning region is larger, or
equivalently if deep material from the complete Si-burning region is
ejected by mixing or aspherical effects.  This can be realized if (i)
the mass cut between the ejecta and the compact remnant is located at
smaller $M_r$ \cite{nakamura1999}, (ii) $E$ is larger to move the outer
edge of the complete Si burning region to larger $M_r$
\cite{nakamura2001b}, or (iii) asphericity in the explosion is larger.

Among these possibilities, a large explosion energy $E$ enhances
$\alpha$-rich freezeout, which results in the increase of the local
mass fractions of Zn and Co, while Cr and Mn are not enhanced
\cite{umeda2002a}.  Models with $E_{51} = 1 $ do not produce
sufficiently large [Zn/Fe].  To be compatible with the observations of
[Zn/Fe] $\sim 0.5$, the explosion energy must be much larger, i.e.,
$E_{51} \gsim 20$ for $M \gsim 20 M_\odot$, i.e., hypernova-like
explosions of massive stars ($M \gsim 25 M_\odot$) with $E_{51} > 10$
are responsible for the production of Zn.

Figure~\ref{fig3} exhibits that the high-energy models tend to be
located at lower [Fe/H] in the SN-induced star formation scenario.
Here we use the HN models with $E_{51}=$ 10, 10, 20, 30, 40 for
$M_{\rm ms} = 20, 25, 30, 40, 50 M_\odot$, respectively.  In this
scenario, EMP stars are enriched by a single supernova \cite{aud95}
which ejects Fe of mass $M$(Fe) and the hydrogen mass swept by
supernova ejecta is proportional to the explosion energy.  Then this
model gives: 
\begin{equation}
{\rm [Fe/H] = log}_{10}(M{\rm (Fe)}/E_{51})-{\rm const}.
\end{equation}

[Ni/Fe] and [Zn/Fe] in our models are in good agreements with the
observations, although [Ni/Fe] is slightly smaller than the
observation.  Ni/Fe is larger if $M_{\rm cut}$(ini) is smaller (i.e.,
the mass cut is deeper) because $^{58}$Ni, a main isotope of Ni, is
mainly synthesized in a deep region with $Y_e < 0.5$.  However, a
smaller $M_{\rm cut}$(ini) tends to suppress Zn/Fe thus requiring
more energetic explosions.

The good agreement of [Zn/Fe] strongly support the SN-induced star
formation model and suggest that EMP stars with lower [Fe/H] might be
made from the ejecta of HNe with larger explosion energies and larger
progenitor's masses.  See, however, \cite{francois2004}.

If VMP stars with [Fe/H] $\sim -2.5$ is made from normal SNe with
$E_{\rm 51}=1$, Zn is underproduced in our models. 1D nucleosynthesis
studies with neutrino transport \cite{frohlich06, frohlich06b, prue05}
suggested that Zn in the normal SN model is enhanced to [Zn/Fe] $\sim
0$ as well as the enhancement of [Sc/Fe].  However, the enhancement is
not large enough to explain the high [Zn/Fe] ($\gsim 0.5$) in the EMP
stars.

Figure~\ref{fig3} shows that [Cr/Fe] in our models is larger than
\cite{cayrel2004}, although the trend in our models is similar to the
observations.  Since Cr is mostly produced in the incomplete
Si-burning region, the relative size of the region should be smaller
than the present model in order to produce smaller Cr/Fe.

[Mn/Fe] and [Co/Fe] in our models are smaller than the observations,
although the trend of [Co/Fe] in our models is similar to the
observations.  Mn can be efficiently enhanced by lowering $Y_e$
\cite{umeda2005} and by a neutrino process \cite{woo95}.  Therefore
the Mn/Fe ratio is important to constrain the physical processes
during the explosion.

\subsection{Carbon-Rich Extremely Metal-Poor Stars}

\begin{figure*}[!ht]
\centering
\includegraphics*[width=12cm]{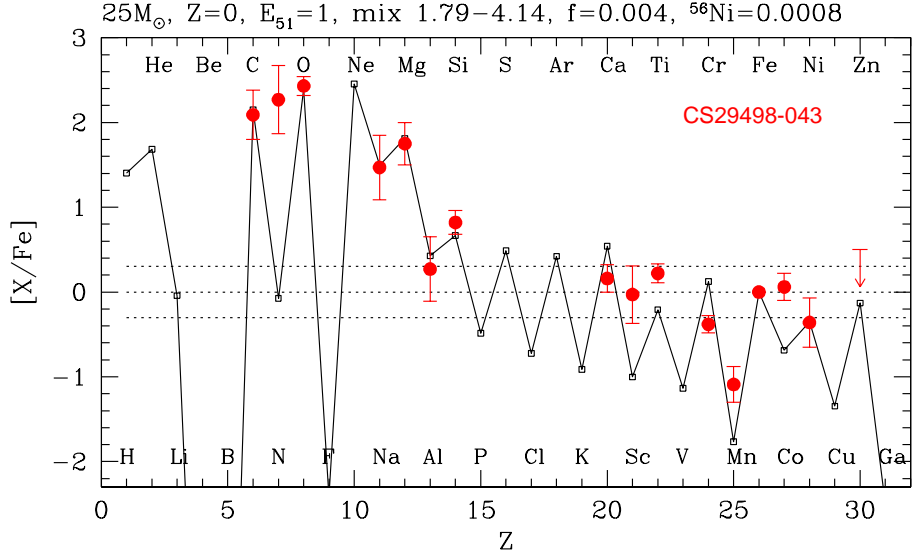}
\caption{
Comparison between the abundance pattern of the C-rich EMP star
(CS~29498-043: {\it filled circles with error bars} \cite{aoki2004})
and the theoretical yields of the 25 $M_\odot$ faint SN ({\it solid
line} \cite{tominaga2006}).
}
\label{fig:faint}
\end{figure*}

Stars with large [C/Fe] ($\sim 1$), called C-rich EMP stars, are
discussed in \cite{umeda2003, umeda2005}.  The origin of those stars
may be different from those of [C/Fe] $\sim 0$ stars.  The large
[C/Fe] ($\gsim 0.5$) can be understood as the faint SN origin, because
the faint SNe are characterized by a large amount of Fe fallback that
leads to large [(C, N, O)/Fe] \cite{umeda2003, umeda2005}.
Figure~\ref{fig:faint} shows the comparison between the abundance
pattern of C-rich EMP stars (CS~29498-043: \cite{aoki2004}) and the 25
$M_\odot$ faint SN model \cite{tominaga2006}.

Most C-rich EMP stars show O/Mg being significantly larger than the
solar ratio.  Faint SNe enhance [O/Fe] more effectively than [Mg/Fe],
because Mg is synthesized in the inner region and thus fallen-back
onto the central remnant more preferentially than O.  (Note that the
abundance determination of O is subject to the uncertain
hydrodynamical (3D) effects \cite{nissen2002}.)

\section{Hyper Metal-Poor (HMP) Stars}

\subsection{HE0107--5240 \& HE1327--2326}

Recently two hyper metal-poor (HMP) stars,
HE0107--5240~\cite{christlieb2002} and HE1327--2326~\cite{frebel2005},
were discovered, whose metallicity Fe/H is smaller than 1/100,000 of
the Sun (i.e., [Fe/H] $< -5$), being more than a factor of 10 smaller
than previously known extremely metal-poor (EMP) stars.  These
discoveries have raised an important question as to whether the
observed low mass ($\sim$ 0.8~$M_\odot$) HMP stars are actually Pop
III stars \cite{weiss2004, suda2004}, or whether these HMP stars are
the second generation stars being formed from gases which were
chemically enriched by a single first generation supernova
(SN)~\cite{umeda2003, limongi2003}.  This is related to the questions
of how the initial mass function depends on the metallicity
\cite[e.g.,][]{schneider2003}.  Thus identifying the origin of these
HMP stars is indispensable to the understanding of the earliest star
formation and chemical enrichment history of the Universe.

\begin{figure*}
\centering
\includegraphics*[width=7.5cm]{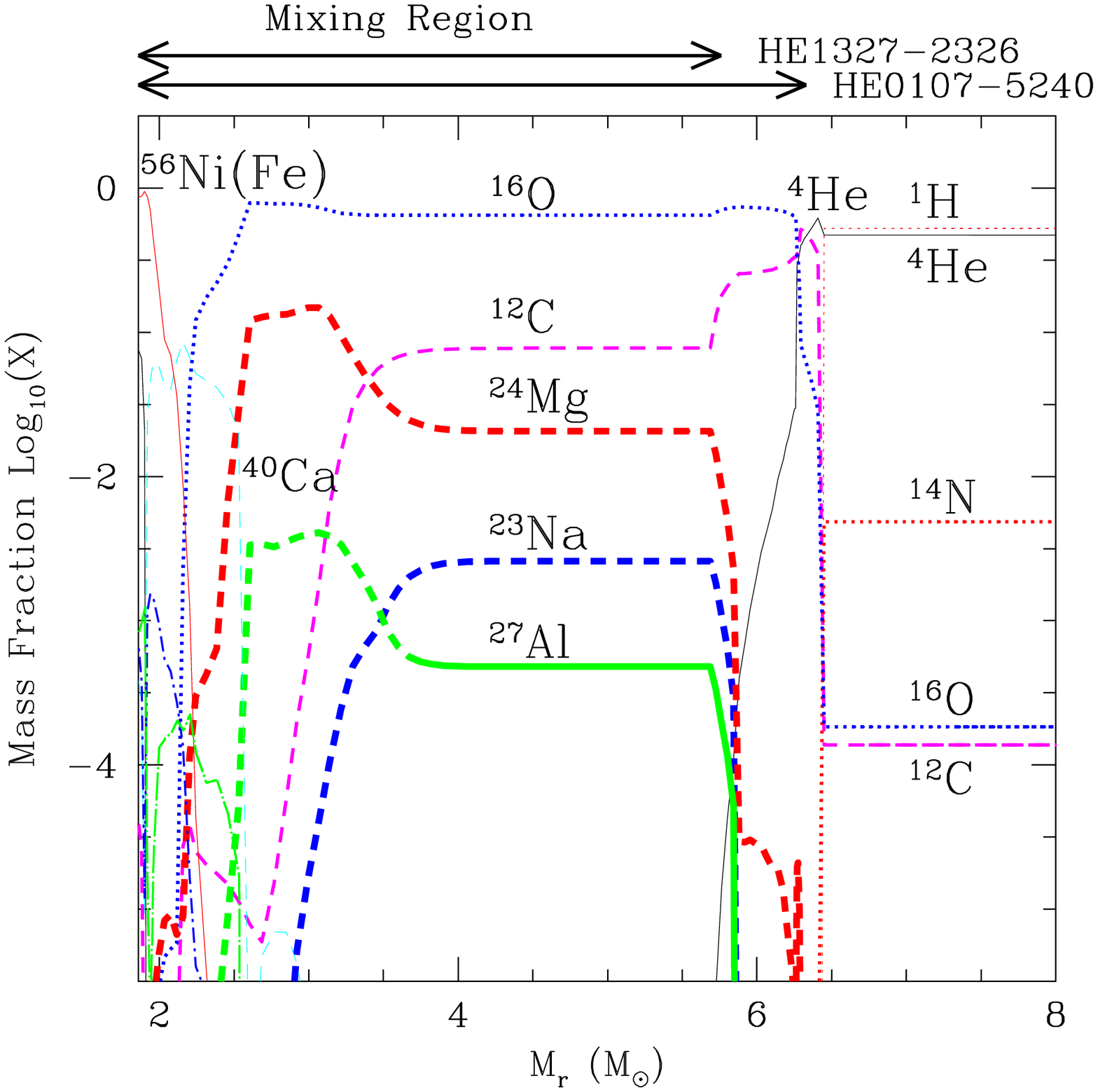}
\includegraphics*[width=6.2cm]{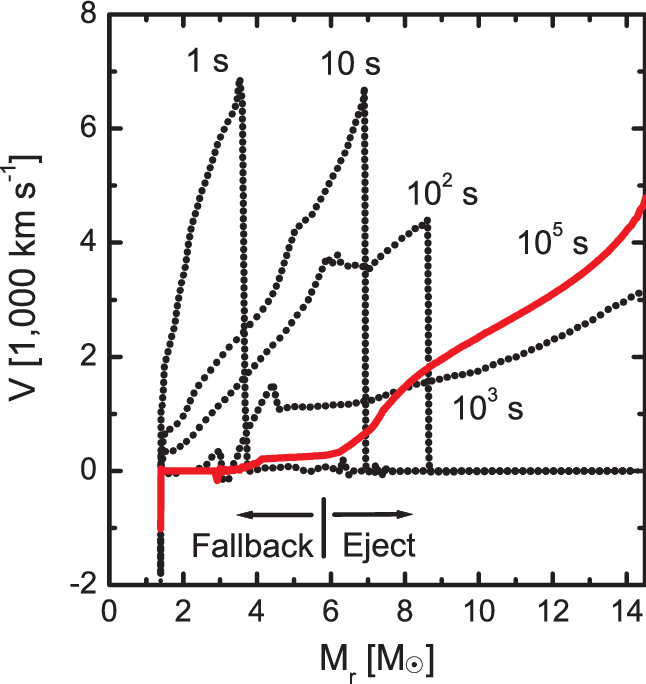}
\includegraphics*[width=12.5cm]{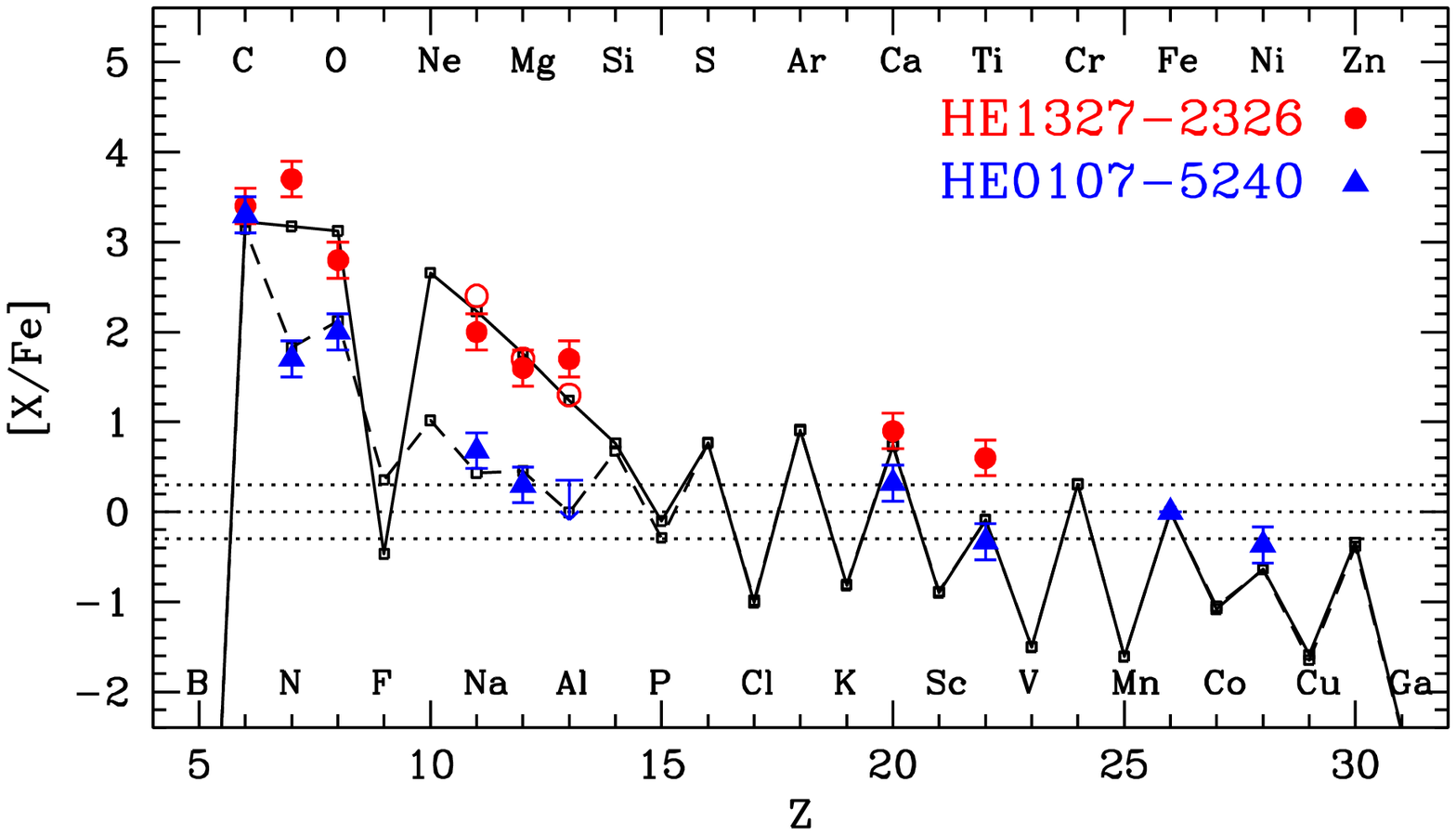}
\caption{(upper-left): The post-explosion abundance distributions for
the 25 $M_\odot$ model with the explosion energy $E_{51} \sim$ 0.7
\cite{iwamoto2005}.  (upper-right): Propagation of the shock wave and
fallback for the HE1327-2326 model \cite{iwamoto2005}.  (lower):
Elemental abundances of the C-rich HMP stars HE0107-5240 \cite[filled
triangles:][]{christlieb2004, christlieb2005} and HE1327--2326
\cite[filled circles:][]{frebel2005, frebel2006} compared with
theoretical supernova yields.}
\label{fig5}
\end{figure*}

 The elemental abundance patterns of these HMP stars provide a key to
the answer to the above questions.  The abundance patterns of
HE1327--2326~\cite{frebel2005, aoki2006, frebel2006} and HE0107--5240
\cite{christlieb2004, bessell2004, christlieb2004} are quite unusual
(Fig. \ref{fig5}). The striking similarity of [Fe/H] (=$-5.4$ and
$-5.2$ for HE1327--2326 and HE0107--5240, respectively) and [C/Fe]
($\sim +4$) suggests that similar chemical enrichment mechanisms
operated in forming these HMP stars.  However, the N/C and (Na, Mg,
Al)/Fe ratios are more than a factor of 10 larger in HE1327--2326.  In
order for the theoretical models to be viable, these similarities and
differences should be explained self-consistently.

Iwamoto et al. \cite{iwamoto2005} showed that the above similarities
and variations of the HMP stars can be well reproduced in unified
manner by nucleosynthesis in the core-collapse ``faint'' supernovae
(SNe) which undergo mixing-and-fallback~\cite{umeda2003}.  We thus
argue that the HMP stars are the second generation low mass stars,
whose formation was induced by the first generation (Pop III) SN with
efficient cooling of carbon-enriched gases.

\subsection{Models for Hyper Metal-Poor Stars}

We consider a model that C-rich EMP stars are produced in the ejecta
of (almost) metal-free supernova mixed with extremely metal-poor
interstellar matter.  The similarity of [Fe/H] and [C/Fe] suggests
that the progenitor's masses of Pop III SNe were similar for these HMP
stars.  We therefore choose the Pop III 25 $M_\odot$ models and
calculate their evolution and explosion \cite{umeda2003, iwamoto2005}.

The abundance distribution after explosive nucleosynthesis is shown in
Figure~\ref{fig5} (upper-left) for the kinetic energy $E$ of the
ejecta $E_{51} \equiv E/10^{51}~{\rm erg} = 0.74$.  The abundance
distribution for $E_{51} = 0.71$ is similar.  In the ``faint'' SN
model, most part of materials that underwent explosive nucleosynthesis
are decelerated by the influence of the gravitational
pull~\cite{woo95} and will eventually fall back onto the central
compact object.  The explosion energies of $E_{51} = 0.74$ and $0.71$
lead to the mass cut $M_{\rm cut} = 5.8 M_\odot$ and $6.3 M_\odot$,
respectively (Fig.\ref{fig5}: upper-right), and the former and the
latter models are used to explain the abundance patterns of
HE1327--2326 and HE0107--5240, respectively.

 During the explosion, the SN ejecta is assumed to undergo mixing,
i.e., materials are first uniformly mixed in the mixing-region
extending from $M_r = 1.9 M_\odot$ to the mass cut at $M_r = M_{\rm
cut}$ (where $M_r$ is the mass coordinate and stands for the mass
interior to the radius $r$) as indicated in Figure~\ref{fig5} (left),
and only a tiny fraction, $f$, of the mixed material is ejected from
the mixing-region together with all materials at $M_r > M_{\rm cut}$;
most materials interior to the mass cut fall back onto the central
compact object.  Such a mixing-fallback mechanism (which might mimic a
jet-like explosion) is required to extract Fe-peak and other heavy
elements from the deep fallback region into the
ejecta~\cite{umeda2003, umeda2005}.

 Figure~\ref{fig5} (lower) shows the calculated abundance ratios in
the SN ejecta models for suitable choice of $f$ which are respectively
compared with the observed abundances of the two HMP stars.  To
reproduce [C/Fe] $\sim$ +4 and other abundance ratios of HMP stars in
Figure~\ref{fig5} (lower), the ejected mass of Fe is only 1.0 $\times
10^{-5}M_\odot$ for HE1327--2326 and 1.4 $\times 10^{-5}M_\odot$ for
HE0107--5240.  These SNe are much fainter in the radioactive tail than
the typical SNe and form massive black holes of $\sim 6 M_\odot$.

 The question is what causes the large difference in the amount of
Na-Mg-Al between the SNe that produced HE0107--5240 and HE1327--2326.
Because very little Na-Mg-Al is ejected from the mixed fallback
materials (i.e., $f \sim 10^{-4}$) compared with the materials
exterior to the final mass cut $M_{\rm cut}$(fin), the ejected amount
of Na-Mg-Al is very sensitive to the location of the mass cut.  As
indicated in Figure~\ref{fig5}, $M_{\rm cut}$(fin) is smaller (i.e.,
the fallback mass is smaller) in the model for HE1327--2326 ($M_{\rm
cut}({\rm fin}) = 5.8 M_\odot$) than HE0107--5240 ($M_{\rm cut}({\rm
fin}) = 6.3 M_\odot$), so that a larger amount of Na-Mg-Al is ejected
from the SN for HE1327--2326.  Since $M_{\rm cut}$(fin) is sensitively
determined by the explosion energy, the (Na-Mg-Al)/Fe ratios among the
HMP stars are predicted to show significant variations and can be used
to constrain $E_{51}$.  Note also that the explosion energies of these
SN models with fallback are not necessarily very small (i.e., $E_{51}
\sim 0.7$).  Further these explosion energies are consistent with
those observed in the actual ``faint'' SNe~\cite{turatto1998}.

 The next question is why HE1327--2326 has a much larger N/C ratio
than HE0107--5240.  In our models, a significant amount of N is
produced by the mixing between the He convective shell and the H-rich
envelope during the presupernova evolution~\cite{umeda2000}, where C
created by the triple-$\alpha$ reaction is burnt into N through the CN
cycle.  For the HE1327--2326 model, we assume about 30 times larger
diffusion coefficients (i.e., faster mixing) for the H and He
convective shells to overcome an inhibiting effect of the mean
molecular weight gradient (and also entropy gradient) between H and He
layers.  Thus, larger amounts of protons are carried into the He
convective shell.  Then [C/N] $\sim 0$ is realized as observed in
HE1327--2326.  Such an enhancement of mixing efficiency has been
suggested to take place in the present-day massive stars known as fast
rotators, which show various N and He enrichments due to different
rotation velocities~\cite{heger2000, meynet2006}.

Recent works have taken into account the NLTE effects and 3D effects,
and obtained [O/Fe] \cite{frebel2006, christlieb2005}.  The resultant
abundance patters are shown in Figure \ref{fig5} (lower) and the newly
obtained [O/Fe] in HE1327--2326 is consistent with our theoretical
models.

\section{Nucleosynthesis in Aspherical Supernovae}

If supernova explosions are not spherically symmetric, resultant
nucleosynthesis is somewhat different from spherical explosions.  Here
we discuss nucleosynthesis in bipolar (jet-like) explosions
\cite{nagataki2000, maeda2002, mae03, maeda2006}.  

In the jet-like explosion model, the shock wave is stronger along the
$z$-axis and heats up the stellar material to higher temperatures.
Along the $r$-axis, temperatures are lower because of the weaker shock
and densities are higher because of mass accretion.  Therefore, the
materials along the $z$-axis occupy the high entropy region in the
$\rho - T$ plane, while those in the r-axis form the lowest bound of
entropy.  As a result, $^{56}$Ni is synthesized preferentially along
the $z$-axis, while a lot of unburned materials, mainly O, are left at
low velocity in the $r$-plane.

\begin{figure}
\centering
\includegraphics[width=6.2cm]{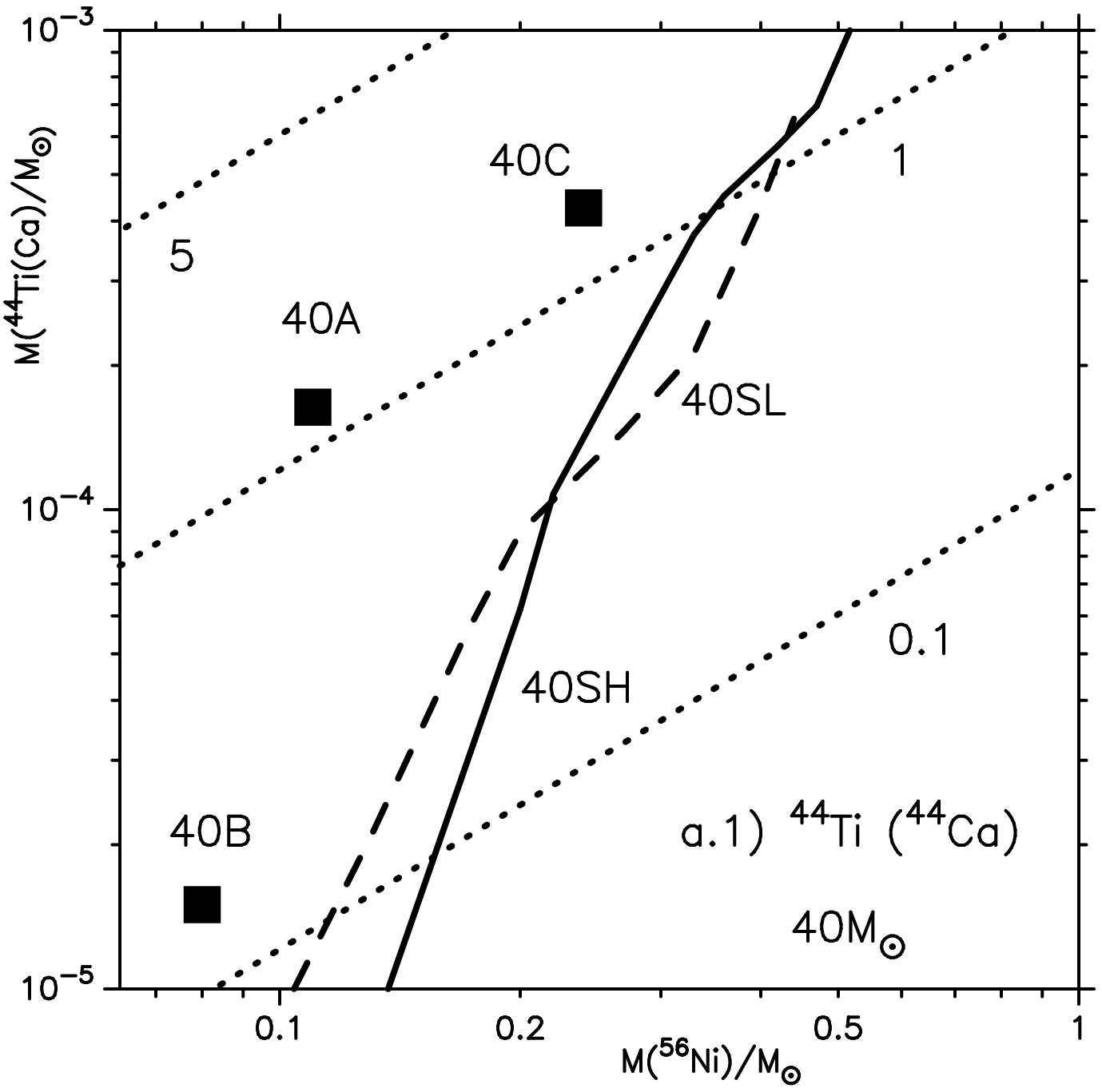}
\includegraphics[width=6.2cm]{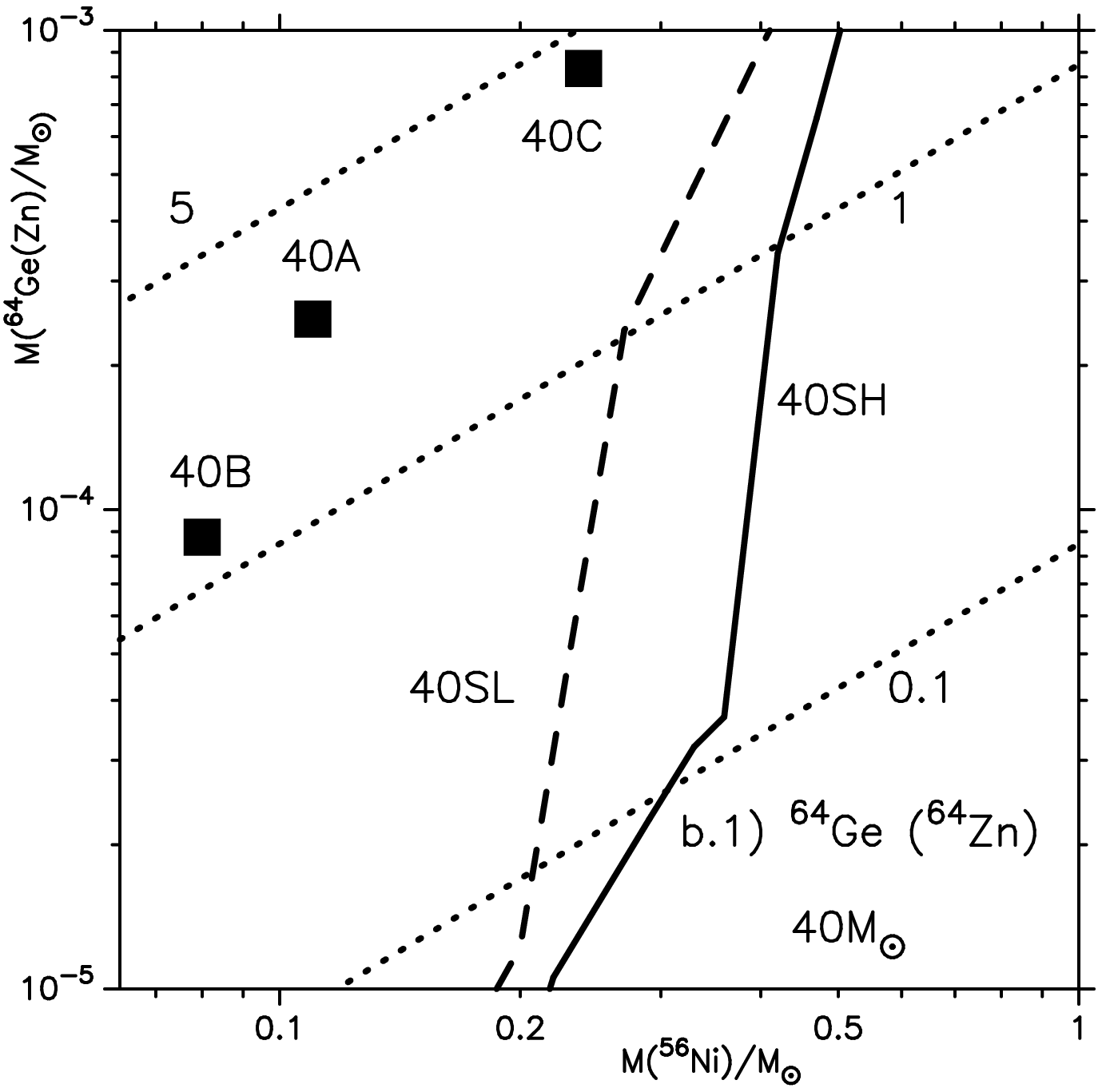}
\caption{Ejected masses of $^{44}$Ti (left) and $^{64}$Ge (right) as a
function of $M$($^{56}$Ni) for some bipolar models (40A, B, and C:
filled squares) and for spherical models (40SH, SL: lines). The dotted
lines show the ratio ($^{44}$Ca, $^{64}$Zn)/$^{56}$Fe relative to the
solar value \cite{mae03}.}
\label{fmae2}
\end{figure}

Because the bipolar models preferentially eject the materials
experiencing higher temperatures (higher entropies) in complete
silicon burning, the ratios $^{44}$Ti/$^{56}$Ni and
$^{64}$Ge/$^{56}$Ni are significantly larger in the bipolar models
than spherical models.  Such an enhancement of the $\alpha$-rich
freezeout products is seen in Figure \ref{fmae2}.  For a given mass of
$^{56}$Ni, the aspherical models (denoted as 40A, B, C) eject larger
amount of $^{44}$Ti and $^{64}$Zn.  Such enhancement could be
important to solve problems related to the production of $^{44}$Ti
\cite{the2006}.

\begin{figure}
\centering
\includegraphics*[width=6.5cm]{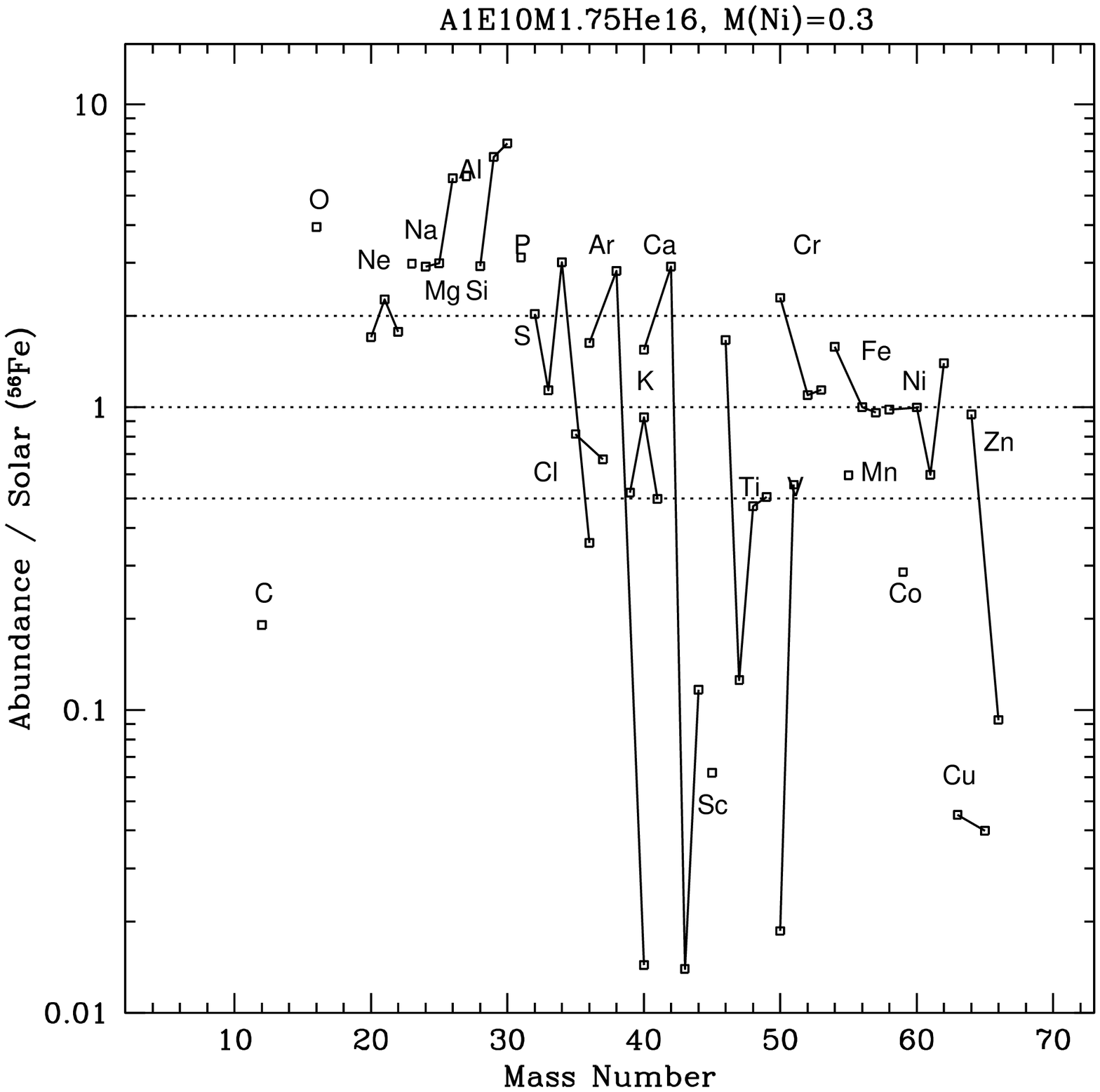}
\includegraphics*[width=6.5cm]{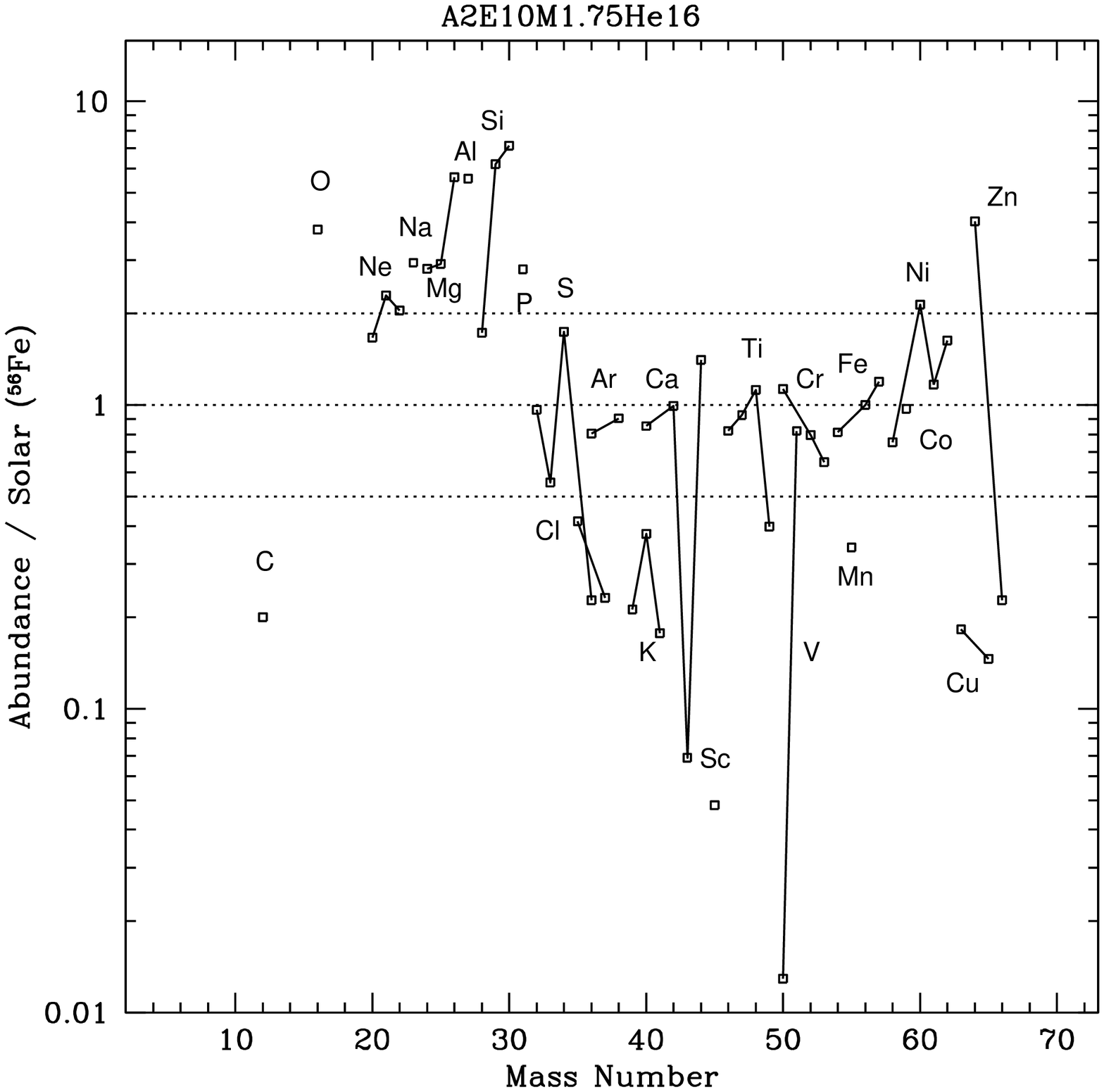}
\caption{Isotopic yields of spherical (left) and aspherical (right)
explosions \cite{maeda2006}.  In both cases, the progenitor is the
$16\Msun$ He core of the $40\Msun$ star, and the explosion energy is
$E_{51} \equiv E/10^{51}$ erg $= 10$.  The mass of $^{56}$Ni is
$0.3\Msun$ (spherical) and $0.24\Msun$ (aspherical), and these values
give [O/Fe] $\sim 0.5$ being consistent with observed in extremely
metal poor stars.}
\label{fmae1}
\end{figure}

As a result, the aspherical models produce the overall abundance
patterns being different from spherical models (Fig. \ref{fmae1}).
[(Zn, Co)/Fe] are enhanced, while [(Mn, Cr)/Fe] are suppressed.  These
trends are similar to ones observed in EMP stars \cite{mcw95}, thus
suggesting important roles of hypernovae in the early Galactic
chemical evolution.

\section{The First Stars}

It is of vital importance to identify the first generation stars in
the Universe, i.e., totally metal-free, Pop III stars.  The impact of
the formation of Pop III stars on the evolution of the Universe
depends on their typical masses.

\subsection{High Mass vs. Low Mass}

Recent numerical models have shown that, the first stars are as
massive as $\gsim$ 100 $M_\odot$ \cite{abel2002, bromm2004} (VMS: very
massive stars).  The formation of long-lived low mass Pop III stars
may be inefficient because of slow cooling of metal free gas cloud,
which is consistent with the failure of attempts to find Pop III
stars.

If the HMP stars are Pop III low mass stars that has gained its metal
from a companion star or interstellar matter \cite{yoshii1981}, would
it mean that the above theoretical arguments are incorrect and that
such low mass Pop III stars have not been discovered only because of
the difficulty in the observations?

Based on the results in the earlier section, we propose that the first
generation supernovae were the explosion of $\sim$ 20-130 $M_\odot$
stars and some of them produced C-rich, Fe-poor ejecta.  Then the low
mass star with even [Fe/H] $< -5$ can form from the gases of mixture
of such a supernova ejecta and the (almost) metal-free interstellar
matter, because the gases can be efficiently cooled by enhanced C and
O ([C/H] $\sim -1$).

\begin{table}[t]

\caption{Stabilities of Pop III and Pop I massive stars: $\bigcirc$
and $\times$ indicate that the star is stable and unstable,
respectively.  The $e$-folding time for the fundamental mode is shown
after $\times$ in units of $10^4$yr \cite{nomoto2003}.}
\begin{center}
\footnotesize
\begin{tabular}{ccccccc}
\hline \hline
 $M(M_\odot)$ & 80 & 100 &120 & 150 & 180 & 300 \\ \hline
 Pop III & $\bigcirc$ & $\bigcirc$ & $\bigcirc$ & $\times$ (9.03) & 
 $\times$ (4.83) & $\times$ (2.15) \\ 
 Pop I & $\bigcirc$ & $\times$ (7.02) & $\times$ (2.35) & 
 $\times$ (1.43) & $\times$ (1.21) & $\times$ (1.71) \\ \hline
\end{tabular}
\end{center}
\label{tab:pop3}
\end{table}

\vspace{0.5cm}

\subsection{Pair Instability SNe vs. Core Collapse SNe}

In contrast to the core-collapse supernovae of 20-130 $M_\odot$ stars,
the observed abundance patterns cannot be explained by the explosions
of more massive, 130 - 300 $M_\odot$ stars. These stars undergo
pair-instability supernovae (PISNe) and are disrupted completely
\cite[e.g.,][]{umeda2002a, heger2002}, which cannot be consistent with
the large C/Fe observed in HMP stars and other C-rich EMP stars.  The
abundance ratios of iron-peak elements ([Zn/Fe] $< -0.8$ and [Co/Fe]
$< -0.2$) in the PISN ejecta (Fig.~\ref{fig7}; \cite{umeda2002a,
heger2002}), cannot explain the large Zn/Fe and Co/Fe in the typical
EMP stars \cite{mcw95, norris2001, cayrel2004} and CS22949-037 either.
Therefore the supernova progenitors that are responsible for the
formation of EMP stars are most likely in the range of $M \sim 20 -
130$ $M_\odot$, but not more massive than 130 $M_\odot$
\cite[also][]{tum04}.  This upper limit mass of the Zero Age Main
Sequence (ZAMS) star depends on the stability of massive stars.

To determine the above upper limit mass, non-adiabatic stabilities of
massive ($80M_\odot$ - $300M_\odot$) Pop III stars have been analyzed
using a radial pulsation code following \cite{ibrahim1981,
baraffe2001, nomoto2003}.  Because CNO elements are absent during the
early stage of their evolution, the CNO cycle does not operate and the
star contracts until temperature rises sufficiently high for the
$3\alpha$ reaction to produce $^{12}$C.  These stars were found to
have $X_{\rm CNO} \sim 1.6 - 4.0 \times10^{-10}$, and the central
temperature $T_{\rm c} \sim 1.4 \times 10^8$ K on their ZAMS.

As summarized in Table~\ref{tab:pop3} \cite{nomoto2003}, the critical
mass of ZAMS Pop III star is $128M_\odot$ while that of Pop I star is
$94M_\odot$.  This difference comes from very compact structures (with
high $T_c$) of Pop III stars.  Stars more massive than the critical
mass will undergo pulsation and mass loss. We note that the
$e$-folding time of instability is much longer for Pop III stars than
Pop I stars with the same mass, and thus the mass loss rate is much
lower. These results are consistent with \cite{ibrahim1981,
baraffe2001}.  The absence of the indication of PISNe in EMP stars
might imply that these massive stars above 130$M_\odot$ undergo
significant mass loss, thus evolving into Fe core-collapse rather than
PISNe.

Alternative possibility is that the First Stars were even more massive
than $\sim 300 M_\odot$ \cite{wasserburg2000, qian2005, fryer2001,
ohkubo2006}.  Such massive stars undergo core-collapse (CVMS:
core-collapse VMS) to form Intermediate mass black holes.  If such
stars formed rapidly rotating black holes, jet-like mass ejection
could produce interesting nucleosynthesis materials to compare with
the elemental abundances observed in the ICM, IGM, and M82
\cite{ohkubo2006}.

\section{Galactic Chemical Evolution}

\subsection{Nucleosynthesis and Metallicity}

\begin{figure*}
\centering
\includegraphics*[width=6.5cm]{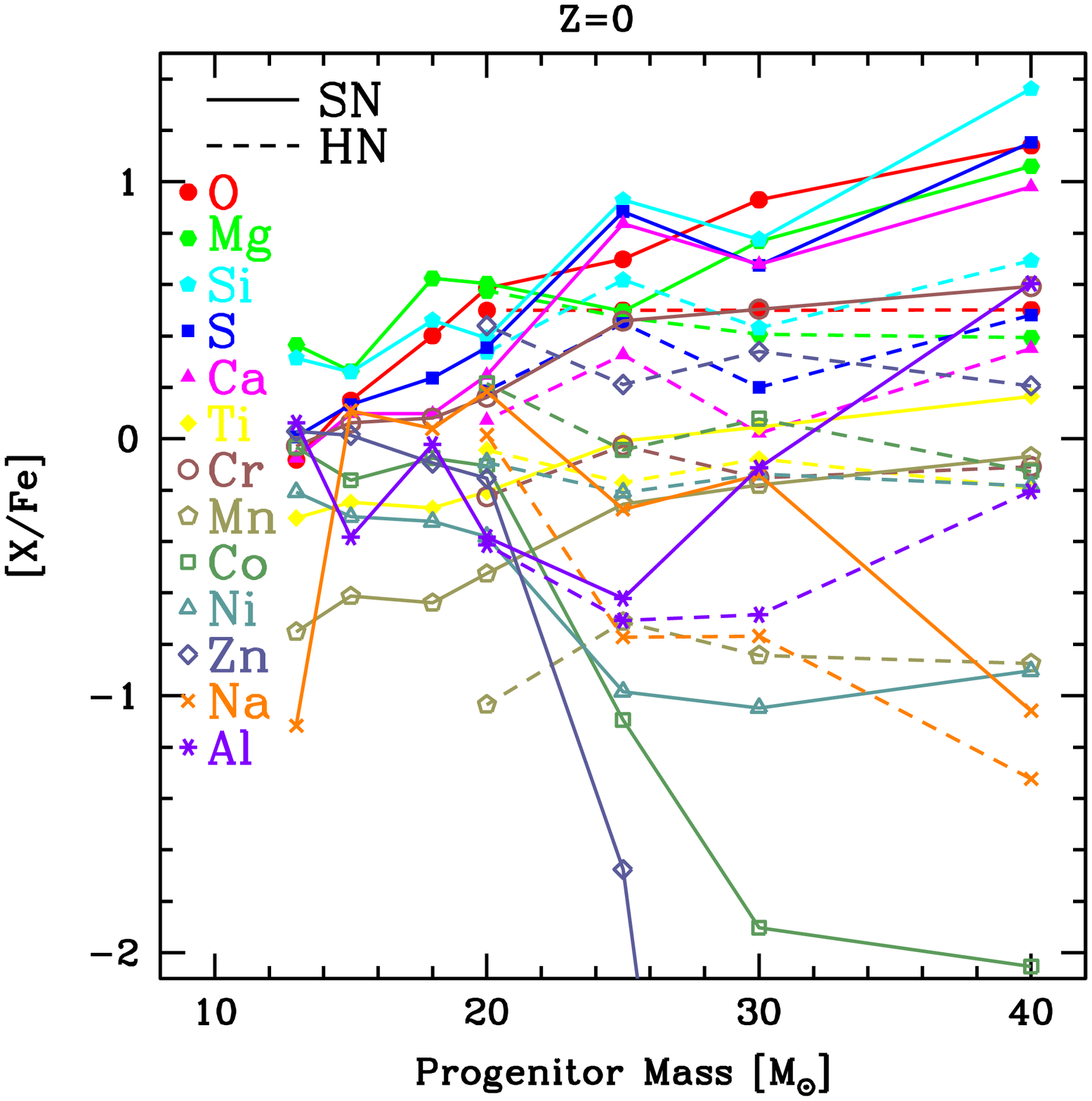}
\includegraphics*[width=6.5cm]{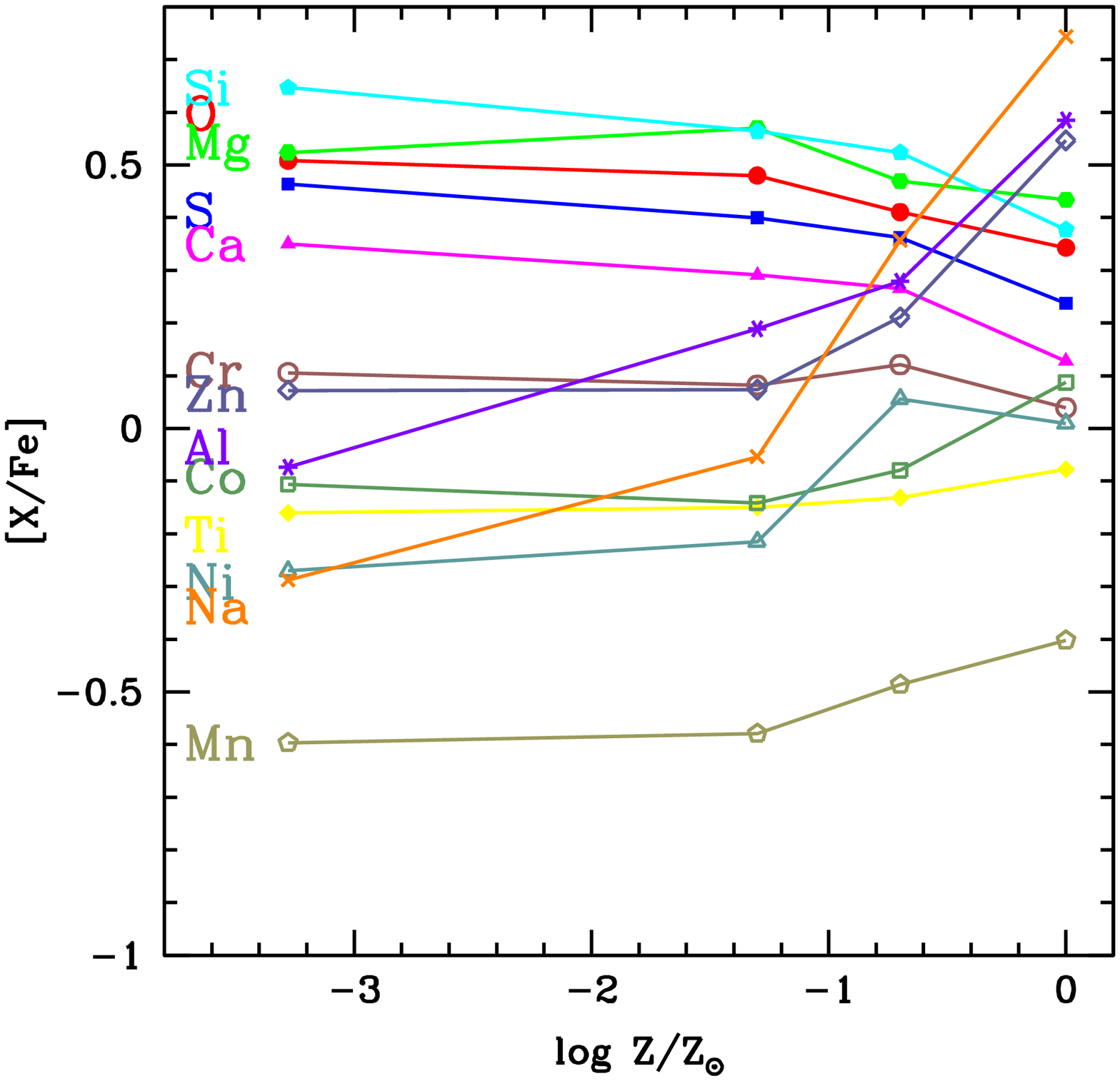}
\caption{(left:) Relative abundance ratios as a function of progenitor 
mass with $Z=0$.  The solid and dashed lines show normal SNe II with
$E_{51}=1$ and HNe.  (right:) The IMF weighted abundance ratios as a
function of metallicity of progenitors, where the HN fraction
$\epsilon_{\rm HN}=0.5$ is adopted.  Results for $Z=0$ are plotted at
$\log Z/Z_\odot=-4$.}
\label{fig:yield}
\end{figure*}

We have calculated the chemical yields for various metallicities,
including Hypernovae \cite{tominaga2006} as described earlier.  For
this set of yields calculations, we assume $Y_{\rm e}=0.4997$ in
incomplete Si-burning region, but use the original $Y_{\rm e}$
\cite{umeda2005} (rather than $Y_{\rm e}=0.5001$) in the complete
Si-burning region.

Table~\ref{tab:chem} gives the resultant nucleosynthesis yields in
units of solar mass for SNe II and HNe as functions of the progenitor
mass ($M =$ 13, 15, 18, 20, 25, 30, 40 $M_\odot$) and metallicity ($Z
=$ 0, 0.001, 0.004, 0.02).  Table~\ref{tab:imf} gives the IMF weighted
yields as a function of metallicity normalized by the total amount of
gases forming stars as follows:
\begin{equation}
 X({\rm A})={\int^{50\Msun}_{0.07\Msun} X_M({\rm A}) M_{\rm ej}(M) M^{-2.35} dM
  \over{\int^{50\Msun}_{0.07\Msun} M \times M^{-2.35} dM}}
\end{equation}
where $X({\rm A})$ is an integrated mass fraction of an element, A,
$X_M({\rm A})$ is mass fraction of A in a model whose mass is nearest
to $M$, and $M_{\rm ej}(M)$ is an ejected mass interpolated between
the nearest models or the nearest model and the edge of the IMF
integrated mass range.  Here we assume $M\leq10M_\odot$ and
$50M_\odot$ stars do not produce materials as Type II SNe or HNe,
i.e., $M_{\rm ej}(M\leq10M_\odot)=M_{\rm ej}(50M_\odot)=0$.

For HNe, we set $E_{51}= 10, 10, 20$, and $30$ for $20$, $25$, $30$,
and $40M_\odot$, respectively.  Although Fe production is larger for
more massive stars because of the higher energy, [$\alpha$/Fe] is
almost constant independent of the stellar mass because we assume the
mass-cut to get [O/Fe] $=0.5$.  In this set, the abundance ratios of
iron-peak elements (Cr, Mn, Co, and Ni) are almost constant because
the mixing-fallback parameters are chosen to give the largest [Zn/Fe].

Figure~\ref{fig:yield} shows the abundance ratios of Pop III ($Z=0$)
SNe II and HNe as a function of the progenitor mass (left) and the IMF
weighted yields of SNe II and HNe as a function of metallicity
(right).  The solid and dashed lines show the SN II and HN yields,
respectively.

The yield masses of $\alpha$ elements (O, Ne, Mg, Si, S, Ar, Ca, and
Ti) are larger for more massive stars because of the larger mantle
mass.  Since the Fe mass is $\sim 0.1 M_\odot$, being independent of
the progenitor's mass for $E_{51}=1$, the abundance ratio
[$\alpha$/Fe] is larger for more massive stars.

In the metal-free stellar evolution, because of the lack of initial
CNO elements, the CNO cycle dose not operate until the star contracts
to a much higher central temperature ($\sim 10^8$ K) than population
II stars, where the 3$\alpha$ reaction produces a tiny fraction of
$^{12}$C ($\sim 10^{-10}$ in mass fraction).  

However, the late core evolution and the resulting Fe core masses of
metal-free stars are not significantly different from metal-rich
stars.  Therefore, [$\alpha$/Fe] is larger by only a factor of $\sim
0.2$ dex and the abundance ratios of the iron-peak elements are not so
different from metal-rich stars, except for Mn.  

On the other hand, the CNO cycle produces only a small amount of
$^{14}$N, which is transformed into $^{22}$Ne during He-burning.  The
surplus of neutrons in $^{22}$Ne increases the abundances of odd-Z
elements (Na, Al, P, ...).  As a result, the abundances of odd-Z
elements depend on the metallicity.  [Na/Fe] and [Al/Fe] of metal-free
stars are smaller by $\sim 1.0$ and $0.7$ dex than solar abundance
stars, which are consistent with the observed trends.

\subsection{Chemical Evolution of the Solar Neighborhood}

\begin{figure*}
\includegraphics*[width=14cm]{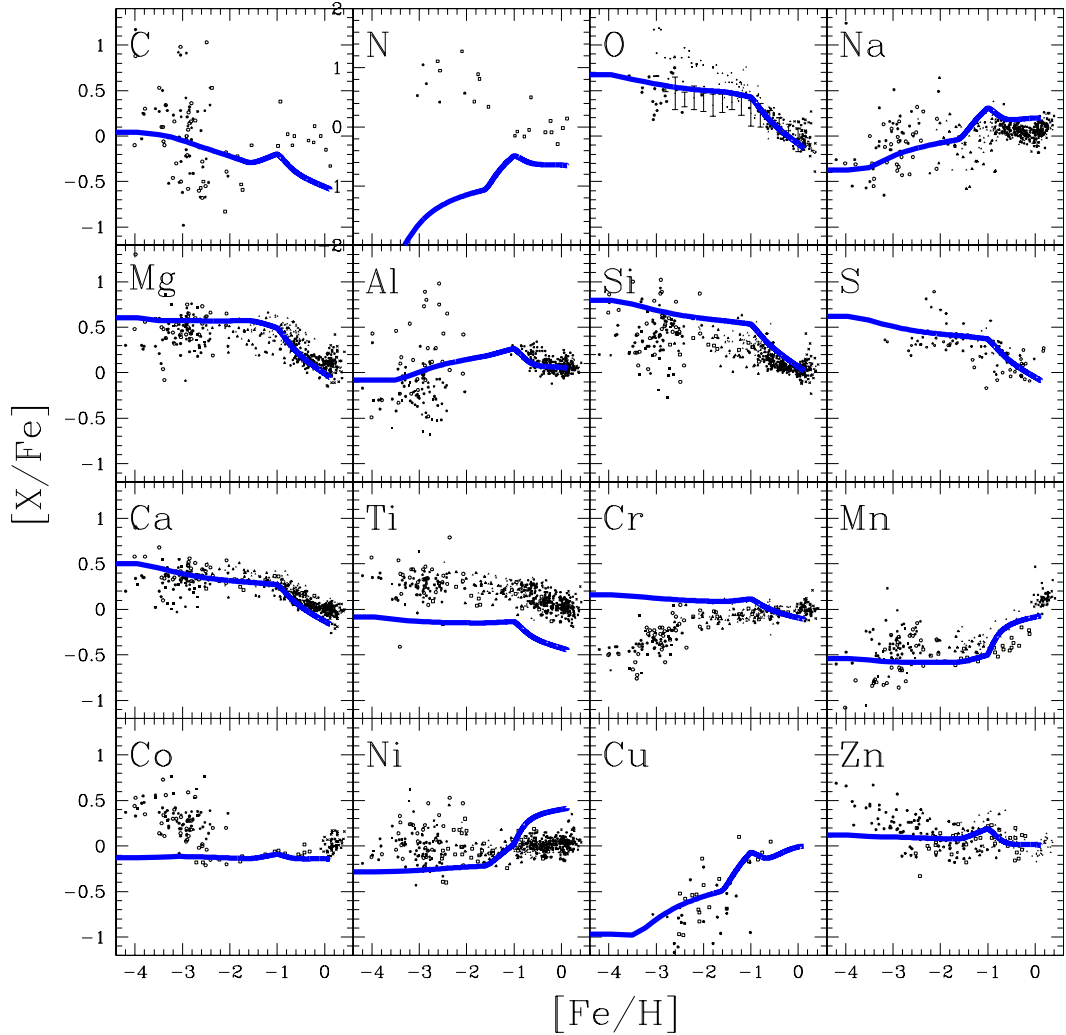}
\caption{
[X/Fe]-[Fe/H] relations for the model with our yields (solid line)
compared with observational
data\cite{ben03,cayrel2004,edv93,gra03,hon04,mcw95,mel02,ryan1996}.
}
\label{fig:xfe}
\end{figure*}

In the chemical evolution model, we should introduce one important
parameter to describe the fraction of hypernovae, $\epsilon_{\rm HN}$.
Although the mass-energy relation has been obtained from the light
curve modeling for individual supernovae, there is no clear constraint
on the energy distribution function because of the poor statistics.
$\epsilon_{\rm HN}$ may depend on metallicity, and may be constrained
with the GRB rate.  Here we adopt $\epsilon_{\rm HN}=0.5$ independent
of the mass and metallicity to find that this provides a good
agreement of [$\alpha$/Fe] plateau against [Fe/H] with observations.
We should note that the plateau value depends on the IMF, specifically
on the slope $x$ and the upper limit $M_{\rm u}$.

We use the chemical evolution model that allows the infall of material
from outside the disk region \cite{koba00}.  We adopt the Galactic age
13 Gyr, infall timescale 5 Gyr, star formation coefficient 0.45
Gyr$^{-1}$, and the Salpeter IMF with a slope of $x=1.35$ in the range
of $0.07M_\odot \leq M \leq 50M_\odot$.  The treatment of SNe Ia is
the same as in \cite{koba00} but with [$b_{\rm RG}=0.02, b_{\rm
MS}=0.04$].  The metallicity inhibition of SNe Ia at [Fe/H] $\leq
-1.1$ is included \cite{koba98}.  These parameter can be determined
uniquely from the metallicity distribution function and the
[O/Fe]-[Fe/H] evolutionary trend at [Fe/H] $\gsim -1$ \cite{pagel1997,
matteucci2001}.

Figures \ref{fig:xfe} shows the evolutions of heavy element abundance
ratios [X/Fe] against [Fe/H].  We note \cite{koba05, koba06}:

{\it Oxygen ---} In our model, [O/Fe] $\sim 0.41$ at [Fe/H] $\sim -1$
and slightly increases to $0.55$ at [Fe/H] $\sim -3$.  The gradual
increase in [O/Fe] with decreasing [Fe/H] is due to the increasing
contributions of more massive, metal-poor SNe II and HNe.  The
metallicity dependence is as small as $0.2$ dex between $Z=Z_\odot$
and $Z=0$. The mass dependence is larger, but such dependence is
weaken because HNe produce more Fe.  From [Fe/H] $\sim -1$, [O/Fe]
decreases quickly due to Fe production by SNe Ia
\cite[e.g.,][]{matteucci2001}.  Further studies of the NLTE and 3D
effects on the O-abundance determination are needed
\cite[e.g.,][]{asplund2005}.

{\it Magnesium ---} Cayrel et al. \cite{cayrel2004} claimed that [(Mg,
Si, Ca, Ti)/Fe] $\sim 0.2-0.3$ being almost constant with a very small
dispersion of $\sim 0.1$ dex.  In our model, [Mg/Fe] $\sim 0.48$ at
[Fe/H] $\sim -1$ and slightly increases to $0.62$ at [Fe/H] $\sim -3$,
which is larger than [Mg/Fe] $\sim 0.27$ in \cite{cayrel2004}, but in
good agreement with the observed relation over the wide range of
[Fe/H].  SNe II with $E_{51}=1$ typically produce [Mg/Fe] $\sim0.5$
with variation from $-0.2$ ($Z=Z_\odot$) to $1$ ($40M_\odot$).

{\it Silicon and Sulfur ---} Observed Si abundance is represented by
only two lines and affected by the contamination.  [Si/Fe] is
$0.52-0.60$ for $-3\lsim$ [Fe/H] $\lsim-1$ in our model, which is
slightly larger than $0.37$ in \cite{cayrel2004} and other
observations.  For S, because of the hardness of observation, the
plateau value is unknown, and our prediction is [S/Fe] $=0.37-0.45$.
Some observations suggest a sharp increase with decreasing [Fe/H]
\cite{isr01}.

{\it Calcium and Titanium ---} Our model succeeds in reproducing the
observation with a plateau [Ca/Fe] $\sim0.31-0.39$.  However, Ti is by
a factor of $\sim 0.4$ dex underabundant overall, which cannot be
improved by changing our parameters.  Possible solution to increase Ti
is a jet-like explosion with high temperature as discussed in \S6
\cite{mae03}.

{\it Sodium, Aluminum, and Copper ---} The NLTE effect for Na and Al
is large for metal-poor stars, and the observational data at [Fe/H]
$\lsim -2$ are shifted by $-0.2$ and $+0.5$, respectively
\cite{asplund2005}.  The abundances of odd-Z elements show a strong
metallicity dependence.  In the one-zone chemical evolution model,
however, the decreasing trend of [(Na,Al,Cu)/Fe] toward lower [Fe/H]
is seen more weakly because of the following mass dependence.
Although the timescale when [Fe/H] reaches $-4$ corresponds to the
lifetime of $\sim 8 M_\odot$, the SFR is peaked at $\sim 8$ Gyr and
stars are forming, and thus more massive stars contribute for lower
metallicity.  [Na/Fe] for $25-30M_\odot, Z=0.001$ and [Al/Fe] for
$40M_\odot, Z=0-0.001$ are as large as $\sim 0.5$, which contribute at
[Fe/H] $\lsim -2.5$.

{\it Chromium, Manganese, Cobalt, and Nickel ---} As seen in Figure
\ref{fig:trend}, the decreasing trend of [(Cr,Mn)/Fe] and the
increasing trend of [Co/Fe] toward lower [Fe/H] are reproduced by
invoking the energy dependence of nucleosynthesis \cite{umeda2005,
tominaga2006} in the SN-induced star formation model. In the
homogeneous one-zone model, the average yields do not show such
trends.
 From [Fe/H] $\sim-1$, SNe Ia contributes to increase [Mn/Fe] toward
higher metallicity.  This is confirmed both observationally and
theoretically, and Mn can be a key element to discuss SNe Ia
contribution, HN fraction, and IMF.

{\it Zinc ---} Zn is an important element observed in damped
Ly$\alpha$ (DLA) systems without the dust depletion effect.  At
$-2\lsim$ [Fe/H] $\lsim-1$, [Zn/Fe] is constant to be $\sim 0.1$, and
mildly decreases to $\sim0$ from [Fe/H] $\sim-1$ due to SNe Ia, which
are consistent with observations.  The increasing trend of [Zn/Fe]
toward lower metallicity \cite{primas00, cayrel2004} cannot be
explained by homogeneously mixed model, but needs to introduce
inhomogeneous models, i.e., SN-induced star formation.

\section{Concluding Remarks}

We have calculated the nucleosynthesis yields for various stellar
masses, explosion energies, and metallicities \cite{tominaga2006}.
 From the light curve and spectra fitting of individual supernova, the
relations between the mass of the progenitor, explosion energy, and
produced $^{56}$Ni mass (witch decayed to $^{56}$Fe) have been
obtained.  Comparison with the abundance patterns of HMP/EMP/VMP stars
has also provided excellent opportunities to test the explosion models
and their nucleosynthesis.

Nucleosynthesis yields of these SNe and HNe are in better agreement
with the observations than before.  In particular, the large Zn
and Co abundances and the small Mn and Cr abundances observed in very
metal-poor stars (Fig.~\ref{fig:trend}) \cite{tominaga2006}) can
better be explained by introducing HNe.  This would imply
that HNe have made significant contributions to the early Galactic
chemical evolution,

In theoretical models, some element ratios, such as (K, Sc, Ti, V)/Fe,
are too small, while some ratios such as Cr/Fe are too large compared
with the observed abundance ratios \cite{cayrel2004}.  Underproduction
of Sc and K may require significantly higher entropy environment for
nucleosynthesis, e.g., the ``low density'' progenitor models for K,
Sc, and Ti \cite{umeda2005}.  GRBs would have possible nucleosynthesis
site, such as accretion disks around the black hole \cite{prue05}.

We need to investigate the uncertainties in the reaction rates which
would be important for synthesis of K and Cr (or ${\rm ^{52}Fe}$ that
decays into ${\rm ^{52}Cr}$).  Rates of some weak processes would need
to be refined in view of critical importance of $Y_e$ near the mass
cut.  In particular, neutrino processes in the deepest layers of SN
ejecta near the mass cut, and accretion disk onto a black hole, would
open a new window for SN nucleosynthesis \cite{prue05, frohlich06,
frohlich06b, wanajo2006}.


This work has been supported in part by the Grant-in-Aid for
Scientific Research (16540229, 17030005, 17033002, 18104003, 18540231)
and the 21st Century COE Program (QUEST) from the JSPS and MEXT of
Japan.

\clearpage

\renewcommand{\arraystretch}{0.42}\selectfont
\begin{table}
\caption{\label{tab:chem}
The yields of individual SN models.}
\begin{tabular}{lccccccc}
\hline
{\tiny $Z=0$}&&&&&&&\\
\hline
{\tiny $M$}&{\tiny 13}&{\tiny 15}&{\tiny 18}&{\tiny 20}&
{\tiny 25}&{\tiny 30}&{\tiny 40}\\
{\tiny $E$}&{\tiny 1}&{\tiny 1}&{\tiny 1}&{\tiny 1}&
{\tiny 1}&{\tiny 1}&{\tiny 1}\\
{\tiny $M_{\rm cut}$}&{\tiny 1.57}&{\tiny 1.48}&{\tiny 1.65}&{\tiny 1.66}&
{\tiny 1.92}&{\tiny 2.07}&{\tiny 2.89}\\
{\tiny p          }  & {\tiny 6.59E+00}  & {\tiny 7.58E+00}  & {\tiny 8.43E+00}  & {\tiny 8.77E+00}  & {\tiny 1.06E+01}  & {\tiny 1.17E+01}  & {\tiny 1.40E+01 }\\
{\tiny d          }  & {\tiny 1.49E$-$16}  & {\tiny 1.69E$-$16}  & {\tiny 1.28E$-$16}  & {\tiny 8.66E$-$17}  & {\tiny 2.02E$-$16}  & {\tiny 1.34E$-$16}  & {\tiny 3.46E$-$16 }\\
{\tiny $^{3}$He   }  & {\tiny 4.12E$-$05}  & {\tiny 4.09E$-$05}  & {\tiny 3.33E$-$05}  & {\tiny 4.76E$-$05}  & {\tiny 2.11E$-$04}  & {\tiny 2.06E$-$04}  & {\tiny 2.56E$-$05 }\\
{\tiny $^{4}$He   }  & {\tiny 4.01E+00}  & {\tiny 4.40E+00}  & {\tiny 5.42E+00}  & {\tiny 5.94E+00}  & {\tiny 8.03E+00}  & {\tiny 9.52E+00}  & {\tiny 1.19E+01 }\\
\hline
\end{tabular}
\end{table}
\begin{table}
\begin{tabular}{lccccccc}
\hline
{\tiny $Z=0.001$}&&&&&&&\\
\hline
{\tiny $M$}&{\tiny 13}&{\tiny 15}&{\tiny 18}&{\tiny 20}&
{\tiny 25}&{\tiny 30}&{\tiny 40}\\
{\tiny $E$}&{\tiny 1}&{\tiny 1}&{\tiny 1}&{\tiny 1}&
{\tiny 1}&{\tiny 1}&{\tiny 1}\\
{\tiny $M_{\rm cut}$}&{\tiny 1.65}&{\tiny 1.53}&{\tiny 1.70}&{\tiny 1.85}&
{\tiny 1.91}&{\tiny 2.06}&{\tiny 3.17}\\
{\tiny p          }  & {\tiny 6.44E+00}  & {\tiny 7.45E+00}  & {\tiny 8.46E+00}  & {\tiny 8.43E+00}  & {\tiny 9.80E+00}  & {\tiny 1.10E+01}  & {\tiny 1.29E+01 }\\
{\tiny d          }  & {\tiny 7.19E$-$15}  & {\tiny 3.38E$-$12}  & {\tiny 3.07E$-$15}  & {\tiny 4.35E$-$16}  & {\tiny 4.46E$-$16}  & {\tiny 5.03E$-$16}  & {\tiny 7.03E$-$14 }\\
{\tiny $^{3}$He   }  & {\tiny 1.43E$-$04}  & {\tiny 1.53E$-$04}  & {\tiny 1.57E$-$04}  & {\tiny 1.60E$-$04}  & {\tiny 1.26E$-$04}  & {\tiny 1.44E$-$04}  & {\tiny 1.21E$-$04 }\\
{\tiny $^{4}$He   }  & {\tiny 3.86E+00}  & {\tiny 5.16E+00}  & {\tiny 6.54E+00}  & {\tiny 5.94E+00}  & {\tiny 6.97E+00}  & {\tiny 8.38E+00}  & {\tiny 1.09E+01 }\\
\hline
\end{tabular}
\end{table}
\begin{table}
\begin{tabular}{lccccccc}
\hline
{\tiny $Z=0.004$}&&&&&&&\\
\hline
{\tiny $M$}&{\tiny 13}&{\tiny 15}&{\tiny 18}&{\tiny 20}&
{\tiny 25}&{\tiny 30}&{\tiny 40}\\
{\tiny $E$}&{\tiny 1}&{\tiny 1}&{\tiny 1}&{\tiny 1}&
{\tiny 1}&{\tiny 1}&{\tiny 1}\\
{\tiny $M_{\rm cut}$}&{\tiny 1.61}&{\tiny 1.50}&{\tiny 1.61}&{\tiny 1.76}&
{\tiny 1.68}&{\tiny 2.56}&{\tiny 2.81}\\
{\tiny p          }  & {\tiny 6.37E+00}  & {\tiny 7.11E+00}  & {\tiny 7.47E+00}  & {\tiny 8.95E+00}  & {\tiny 1.02E+01}  & {\tiny 1.01E+01}  & {\tiny 1.03E+01 }\\
{\tiny d          }  & {\tiny 1.07E$-$14}  & {\tiny 1.99E$-$14}  & {\tiny 4.57E$-$14}  & {\tiny 8.80E$-$14}  & {\tiny 3.23E$-$16}  & {\tiny 6.93E$-$14}  & {\tiny 2.64E$-$16 }\\
{\tiny $^{3}$He   }  & {\tiny 1.70E$-$04}  & {\tiny 1.59E$-$04}  & {\tiny 2.24E$-$04}  & {\tiny 1.75E$-$04}  & {\tiny 1.85E$-$04}  & {\tiny 1.84E$-$04}  & {\tiny 1.80E$-$04 }\\
{\tiny $^{4}$He   }  & {\tiny 4.04E+00}  & {\tiny 4.95E+00}  & {\tiny 6.06E+00}  & {\tiny 7.03E+00}  & {\tiny 8.48E+00}  & {\tiny 7.92E+00}  & {\tiny 8.12E+00 }\\
\hline
\end{tabular}
\end{table}
\begin{table}
\begin{tabular}{lccccccc}
\hline
{\tiny $Z=0.02$}&&&&&&&\\
\hline
{\tiny $M$}&{\tiny 13}&{\tiny 15}&{\tiny 18}&{\tiny 20}&
{\tiny 25}&{\tiny 30}&{\tiny 40}\\
{\tiny $E$}&{\tiny 1}&{\tiny 1}&{\tiny 1}&{\tiny 1}&
{\tiny 1}&{\tiny 1}&{\tiny 1}\\
{\tiny $M_{\rm cut}$}&{\tiny 1.60}&{\tiny 1.50}&{\tiny 1.58}&{\tiny 1.55}&
{\tiny 1.69}&{\tiny 2.10}&{\tiny 2.21}\\
{\tiny p          }  & {\tiny 6.15E+00}  & {\tiny 6.79E+00}  & {\tiny 7.53E+00}  & {\tiny 7.93E+00}  & {\tiny 8.41E+00}  & {\tiny 8.75E+00}  & {\tiny 3.55E+00 }\\
{\tiny d          }  & {\tiny 8.34E$-$15}  & {\tiny 1.02E$-$14}  & {\tiny 4.16E$-$16}  & {\tiny 1.20E$-$15}  & {\tiny 9.87E$-$16}  & {\tiny 1.08E$-$15}  & {\tiny 7.71E$-$17 }\\
{\tiny $^{3}$He   }  & {\tiny 1.96E$-$04}  & {\tiny 2.18E$-$04}  & {\tiny 2.30E$-$04}  & {\tiny 2.38E$-$04}  & {\tiny 2.21E$-$04}  & {\tiny 2.12E$-$04}  & {\tiny 5.05E$-$05 }\\
{\tiny $^{4}$He   }  & {\tiny 4.30E+00}  & {\tiny 5.25E+00}  & {\tiny 6.11E+00}  & {\tiny 6.76E+00}  & {\tiny 7.24E+00}  & {\tiny 8.36E+00}  & {\tiny 4.71E+00 }\\
\hline
\end{tabular}%
\end{table}

\begin{table}
\caption{\label{tab:imf}
The IMF weighted yields with $x=1.35, M_{\ell}=0.07M_\odot, M_{\rm u}=50M_\odot$.}
\begin{tabular}{lccccccc}
\hline
	  &\multicolumn{4}{c}{{\tiny SNII+HN}   }\\
{\tiny $Z$       }&{\tiny          0}&{\tiny     0.001}&{\tiny      0.004}&{\tiny       0.02}\\ \hline\hline
{\tiny p         }&{\tiny   3.28E-02}&{\tiny   3.14E-02}&{\tiny   2.96E-02}&{\tiny   2.45E-02}\\
{\tiny d         }&{\tiny   5.76E-18}&{\tiny   2.21E-15}&{\tiny   1.97E-16}&{\tiny   5.34E-16}\\
{\tiny $^3$He    }&{\tiny   2.51E-07}&{\tiny   5.24E-07}&{\tiny   6.52E-07}&{\tiny   7.06E-07}\\
{\tiny $^4$He    }&{\tiny   2.30E-02}&{\tiny   2.26E-02}&{\tiny   2.22E-02}&{\tiny   2.05E-02}\\
{\tiny $^6$Li    }&{\tiny   1.70E-20}&{\tiny   1.52E-19}&{\tiny   6.49E-19}&{\tiny   1.76E-18}\\
{\tiny $^7$Li    }&{\tiny   8.67E-12}&{\tiny   6.67E-13}&{\tiny   3.14E-15}&{\tiny   5.85E-14}\\
\hline
\end{tabular}
\end{table}



\clearpage

\begin{thebibliography}{999}

\bibitem{abel2002} 
Abel, T., Bryan, G.L., \& Norman, M.L., Science {\bf 295} (2002) 93

\bibitem{aoki2004} 
Aoki, W., Norris, J. E., Ryan, S. G., Beers, T. C., Christlieb, N.,
Tsangarides, S., \& Ando, H., \apj {\bf 608} (2004) 971

\bibitem{aoki2006} 
Aoki, W., Frebel, A., Christlieb, N., et al., \apj {\bf 639} (2006)
897

\bibitem{argast2000} 
Argast, D., Samland, M., Gerhard, O.E., \& Thielemann, F.-K., \aa {\bf
356} (2000) 873

\bibitem{arnett1996} 
Arnett, W.D., {\em Supernovae and Nucleosynthesis} (Princeton
Univ. Press) (1996)

\bibitem{asplund2005} 
Asplund, M. \annrev {\bf 43} (2005) 481

\bibitem{aud95} 
Audouze, J., \& Silk, J., \apj {\bf 451} (1995) L49

\bibitem{baraffe2001}
Baraffe, I., Heger, A., \& Woosley, S.E., \apj {\bf 550} (2001) 890

\bibitem{beers2005} 
Beers, T., \& Christlieb, N., \annrev {\bf 43} (2005) 531

\bibitem{ben03}
Bensby, T., Feltzing, S., \& Lundstrom, I., \aa {\bf 410} (2003) 527

\bibitem{bessell2004}
Bessell, M.S., Christlieb, N., \& Gustafsson, B., \apj {\bf 612}
(2004) L61

\bibitem{bromm2004} 
Bromm, V., \& Larson, R., \annrev {\bf 42} (2004) 79

\bibitem{burrows2006}
Burrows, A., Livne, E., Dessart, L., Ott, C.D., \& Murphy, J., \apj
{\bf 640} (2006) 878

\bibitem{campana2006} 
Campana, S., et al., \nat (2006) submitted (astro-ph/0603279)

\bibitem{cayrel2004}
Cayrel, R., et al., \aa {\bf 416} (2004) 1117

\bibitem{cl2002} 
Chieffi, A., \& Limongi, M., \apj {\bf 577} (2002) 281

\bibitem{christlieb2002} 
Christlieb, N., et al., \nat {\bf 419} (2002) 904

\bibitem{christlieb2004} 
Christlieb, N., et al., \apj {\bf 603} (2004) 708

\bibitem{christlieb2005} 
Christlieb, N., in {\em ESO/MPA workshop, Carbon-rich Ultra Metal-Poor
Stars of the Galactic Halo}, ed. A. Weiss \& F. Primas (MPA) (2005)

\bibitem{edv93} 
Edvardsson, B., Andersen, J., Gustafsson, B., Lambert, D.L., Nissen,
P.E., \& Tomkin, J., \aa {\bf 275} (1993) 101

\bibitem{alex1997} 
Filippenko, A.V., \annrev {\bf 35} (1997) 309

\bibitem{francois2004} 
Francois, P., Matteucci, F., Cayrel, R., Spite, M., Spite, F., \&
Chaippini, C., \aa {\bf 421} (2004) 613

\bibitem{frebel2005} 
Frebel, A., et al., \nat {\bf 434} (2005) 871

\bibitem{frebel2006} 
Frebel, A., Christlieb, N., Norris, J.E., Aoki, W., \& Asplund, M.,
\apj {\bf 638} (2006) L17

\bibitem{frohlich06} 
Fr\"ohlich, C., Hauser, P., Liebend\"orfer, M., Mart\'inez-Pinedo, G.,
Thielemann, F.-K., Bravo, E., Zinner, N. T., Hix, W. R., Langanke, K.,
Mezzacappa, A., \& Nomoto, K., \apj {\bf 637} (2006) 415

\bibitem{frohlich06b} 
Fr\"ohlich, C., Mart\'inez-Pinedo, G., Liebend\"orfer, M., Thielemann,
F.-K., et al., {\em Phys Rev Let} {\bf 96} (2006) 142502

\bibitem{fryer2001} 
Fryer, C.L., Woosley, S. E., \& Heger, A., \apj {\bf 550} (2001) 372

\bibitem{fryer2004} 
Fryer, C.L. (ed.) {\em Stellar Collapse} (Astrophysics and Space
Science Library: Kluwer) (2004)

\bibitem{matteucci2002}
Fusco-Femiano, R., \& Matteucci, F. (ed.), {\em ASP Series} {\bf 253}
{\em Chemical Enrichment of Intracluster and Intergalactic Medium},
(ASP) (2002)

\bibitem{galama1998} 
Galama, T., et al., \nat {\bf 395} (1998) 670

\bibitem{gra03}
Gratton, R.G., et al., \aa {\bf 404} (2003) 187

\bibitem{hac90} 
Hachisu, I., Matsuda, T., Nomoto, K., et al., \apj {\bf 358} (1990)
L57

\bibitem{hamuy2003} 
Hamuy, M., \apj {\bf 582} (2003) 905

\bibitem{heger2000} 
Heger, A., \& Langer, N., \apj {\bf 544} (2000) 1016
 
\bibitem{heger2002} 
Heger, A., \& Woosley, S.E., \apj {\bf 567} (2002) 532
 
\bibitem{hill2005}
Hill, V., Francois, P., \& Primas, F. (ed.), {\em IAU Symp 228, From
Lithium to Uranium} (Cambridge Univ. Press) (2005)

\bibitem{hillebrandt2003}
Hillebrandt, W., \& Leibundgut, B. (ed.), {\em From Twilight to
Highlight: The Physics of Supernovae} (Springer) (2003)

\bibitem{hjorth2003} 
Hjorth, J., et al., \nat {\bf 423} (2003) 847
 
\bibitem{hon04}
Honda, S. et al., \apj {\bf 607} (2004) 474

\bibitem{ibrahim1981}
Ibrahim, A., Boury, A., \& Noels, A., \aa {\bf 103} (1981) 390

\bibitem{isr01}
Israelian, G. \& Reboro, R., \apj {\bf 557} (2001) L43

\bibitem{iwamoto1998} 
Iwamoto, K., Mazzali, P.A., Nomoto, K., et al., \nat {\bf 395} (1998) 672

\bibitem{iwa99}
Iwamoto, K, Brachwitz, F, Nomoto, K., Kishimoto, N., Umeda, H., Hix,
W. R., \& Thielemann, F-K., \apjs {\bf 125} (1999) 439

\bibitem{iwamoto2000} 
Iwamoto, K., Nakamura, T., Nomoto, K., et al., \apj {\bf 534} (2000) 660

\bibitem{iwamoto2005} 
Iwamoto, N., Umeda, H., Tominaga, N., Nomoto, K., \& Maeda, K.,
Science {\bf 309} (2005) 451

\bibitem{janka2001} 
Janka, H.-Th. \aa {\bf 368} (2001) 527

\bibitem{kawabata2002} 
Kawabata, K., et al., \apj {\bf 580} (2002) L39

\bibitem{kif03} 
Kifonidis, K., Plewa, T., Janka, H.-Th., et al., \aa {\bf 408} (2003)
621

\bibitem{koba98}
Kobayashi, C., Tsujimoto, T., Nomoto, K., Hachisu, I, \& Kato, M.,
\apj {\bf 503} (1998) L155

\bibitem{koba00} 
Kobayashi, C., Tsujimoto, T., \& Nomoto, K., \apj {\bf 539} (2000) 26

\bibitem{koba05}
Kobayashi, C., in {\em IAU Symp 228, From Lithium to Uranium},
ed. V. Hill et al.  (Cambridge Univ. Press) (2005) 297
(astro-ph/0508000)

\bibitem{koba06}
Kobayashi, C., Umeda, H., Nomoto, K., Tominaga, N., \& Ohkubo, T., 
\apj (2006) submitted

\bibitem{langer1992} 
Langer, N., \aa {\bf 265} (1992) L17

\bibitem{liebend03} 
Liebend\"orfer, M., Mezzacappa, A., Messer, O. E. B., Martinez-Pinedo,
G., Hix, W. R., Thielemann, F.-K., \nphys {\bf A719} (2003) 144

\bibitem{liebend05} 
Liebend\"orfer, M., Rampp, M., Janka, H.-Th., \& Mezzacappa, A., \apj
{\bf 620} (2005) 840

\bibitem{lc2003} 
Limongi, M., \& Chieffi, A., \apj {\bf 592} (2003) 404

\bibitem{limongi2003} 
Limongi, M., Chieffi, A., \& Bonifacio, P., \apj {\bf 594} (2003) L123

\bibitem{maeda2002}
Maeda, K., Nakamura, T., Nomoto, K., Mazzali, P.A., Patat, F., \&
Hachisu, I. \apj {\bf 565} (2002) 405
 
\bibitem{mae03}
Maeda, K. \& Nomoto, K., \apj {\bf 598} (2003) 1163

\bibitem{maeda2006}
Maeda, K., in {\em Origin of Matter and Evolution of Galaxies} (2006)
ed. S. Kubono et al. (AIP), in press

\bibitem{maeder2000} 
Maeder, A. \& Meynet, G., \annrev {\bf 38} (2000) 143

\bibitem{mal04} 
Malesani, J., et al., \apj {\bf 609} (2006) L5

\bibitem{matteucci2001} 
Matteucci, F., {\em The Chemical Evolution of the Galaxy} (Kluwer
Academic Pub.) (2001)

\bibitem{mazzali2000} 
Mazzali, P.A., Iwamoto, K., Nomoto, K., \apj {\bf 545} (2000) 407

\bibitem{mazzali2002} 
Mazzali, P.A., Deng, J., Maeda, K., Nomoto, K., et al., \apj {\bf 572}
(2002) L61

\bibitem{mazzali2003} 
Mazzali, P.A., et al., \apj {\bf 599} (2003) L95

\bibitem{mazzali2006} 
Mazzali, P.A., Deng, J., Nomoto, K., et al., \nat (2006) submitted
(astro-ph/0603567)

\bibitem{mcw95} 
McWilliam, A., Preston, G.W., Sneden, C., et al., \aj {\bf 109} (1995)
2757

\bibitem{mel02}
Melendez, J. \& Barbuy, B., \apj {\bf 575} (2002) 474

\bibitem{meynet2006} 
Meynet, G., Ekstrom, S., \& Maeder, A., \aa {\bf 447} (2006) 623

\bibitem{nagataki2000}
Nagataki, S., \apjs {\bf 127} (2000) 141

\bibitem{nakamura1999} 
Nakamura, T., Umeda, H., Nomoto, K., Thielemann, F.-K., \& Burrows,
A., \apj {\bf 517} (1999) 193

\bibitem{nakamura2001a} 
Nakamura, T., Mazzali, P.A., Nomoto, K., Iwamoto, K., \apj {\bf 550}
(2001a) 991

\bibitem{nakamura2001b} 
Nakamura, T., Umeda, H., Iwamoto, K., Nomoto, K., et al., \apj {\bf
555} (2001b) 880

\bibitem{nissen2002} 
Nissen, P.E., Primas, F., Asplund, M., \& Lambert, D.L., \aa {\bf 390}
(2002) 235

\bibitem{nomoto1994a} 
Nomoto, K., Yamaoka, H., Shigeyama, T., Kumagai, S., \& Tsujimoto, T.,
in {\em Supernovae}, Les Houche Session LIV (1994) ed. S. Bludmann et
al. (North-Holland) 199

\bibitem{nomoto1994b} 
Nomoto, K., Shigeyama, T., Kumagai, S., Yamaoka, H., \& Suzuki, T., in
{\em Supernovae}, Les Houche Session LIV (1994) ed. S. Bludmann et
al. (North-Holland) 489

\bibitem{nomoto1994c} 
Nomoto, K., Yamaoka, H., Pols, O.R., van den Heuvel, E.P.J., et al.,
\nat {\bf 371} (1994) 227

\bibitem{nomoto1997} 
Nomoto, K., Hashimoto, M., Tsujimoto, T., Thielemann, F.-K., et al.,
\nphys {\bf A616} (1997) 79c

\bibitem{nomoto2001} 
Nomoto, K., Mazzali, P.A., Nakamura, T., et al., in {\em Supernovae
and Gamma Ray Bursts}, eds. M. Livio et al. (Cambridge Univ. Press)
(2001) 144 (astro-ph/0003077)

\bibitem{nomoto2003} 
Nomoto, K., Maeda, K., Umeda, H., et al., in {\em IAU Symp 212, A
Massive Star Odyssey, from Main Sequence to Supernova},
eds. V.D. Hucht, et al. (San Francisco: ASP) (2003) 395
(astro-ph/0209064)

\bibitem{nomoto2004} 
Nomoto, K., Maeda, K., Mazzali, P.A., et al., in {\em Stellar
Collapse}, ed. C.L. Fryer (Astrophysics and Space Science Library:
Kluwer) (2004) 277 (astro-ph/0308136)

\bibitem{nomoto2005} 
Nomoto, K., Tominaga, N., Umeda, H., \& Kobayashi, C., in {\em IAU
Symp 228, From Lithium to Uranium}, ed. V. Hill et al. (Cambridge
Univ. Press) (2005) 287 (astro-ph/0603433)

\bibitem{norris2001} 
Norris, J.E., Ryan, S.G., \& Beers, T.C., \apj {\bf 561} (2001) 1034

\bibitem{ohkubo2006} 
Ohkubo, T., Umeda, H., Maeda, K., Nomoto, K., Suzuki, T., Tsuruta, S.,
\& Rees, M.J., \apj {\bf 654} (2006) in press (astro-ph/0507593)

\bibitem{pac1998} 
Paczy\'nski, B., \apj {\bf 494} (1998) L45

\bibitem{pagel1997} 
Pagel, B.E.J., {\em Nucleosynthesis and Chemical Evolution of
Galaxies} (Cambridge Unv. Press) (1997)

\bibitem{pian2006} 
Pian, E., et al., \nat (2006) submitted (astro-ph/0603530)

\bibitem{primas00}
Primas, F., et al., in {\em The First Stars} (2000), ed. A. Weiss,
T. Abel, \&, V. Hill (Berlin: Springer), 51

\bibitem{prue05} 
Pruet, J., Woosley, S.E., Buras, R., Janka, H.-T., \& Hoffman, R.D.,
\apj {\bf 623} (2005) 325

\bibitem{qian2005} 
Qian, Y.-Z., \& Wasserburg, G.J., \apj {\bf 635} (2005) 845

\bibitem{rampp2000} 
Rampp, M., \& Janka, H.-Th., \apj {\bf 539} (2000) L33

\bibitem{rauscher2002} 
Rauscher, T., Heger, A., Hoffman, R.D., \& Woosley, S.E., \apj {\bf
576} (2002) 323
 
\bibitem{ryan1996} 
Ryan, S.G., Norris, J.E., \& Beers, T.C., \apj {\bf 471} (1996) 254

\bibitem{schneider2003} 
Schneider, R., Ferrara, A., Salvaterra, R., Omukai, K., \& Bromm, V.,
\nat {\bf 422} (2003) 869

\bibitem{shigeyama1998} 
Shigeyama, T., \& Tsujimoto, T., \apj {\bf 507} (1998) L135

\bibitem{sneden1991} 
Sneden, C., Gratton, R.G., \& Crocker, D.A., \aa {\bf 246} (1991) 354

\bibitem{sollerman1998} 
Sollerman, J., Cumming, R., \& Lundqvist, P., \apj {\bf 493} (1998)
933

\bibitem{stanek2003} 
Stanek, K.Z., et al., \apj {\bf 591} (2003) L17

\bibitem{suda2004} 
Suda, T., Aikawa, M., Machida, M.N., Fujimoto, M.Y., \& Iben, I.,
Jr., \apj {\bf 611} (2004) 476 

\bibitem{the2006}
The, L.-S., Clayton, D.D., Diehl, R., et al., \aa {\bf 450} (2006)
1037

\bibitem{tnh96} 
Thielemann, F.-K., Nomoto, K., \& Hashimoto, M., \apj {\bf 460} (1996)
408

\bibitem{tominaga2005} 
Tominaga, N., Tanaka, M., Nomoto, K., et al., \apj {\bf 633} (2005)
L97

\bibitem{tominaga2006} 
Tominaga, N., Umeda, H., \& Nomoto, K. \apj (2006) submitted

\bibitem{tum06} 
Tumlinson, J., \apj {\bf 641} (2006) 1

\bibitem{tum04} 
Tumlinson, J., Venkatesan, A., \& Shull, M., \apj {\bf 612} (2004) 602

\bibitem{turatto1998} 
Turatto, M., Mazzali, P.A., Young, T., Nomoto, K., et al., \apj {\bf
498} (1998) L129

\bibitem{umeda2000} 
Umeda, H.  Nomoto, K., \& Nakamura, T., in {\em The First Stars}
(2000), ed. A. Weiss, T. Abel, \& V. Hill (Berlin: Springer), 150
(astro-ph/9912248)

\bibitem{umeda2002a} 
Umeda, H., \& Nomoto, K., \apj {\bf 565} (2002) 385

\bibitem{umeda2002b} 
Umeda, H., Nomoto, K., Tsuru, T., et al., \apj {\bf 578}
(2002) 855

\bibitem{umeda2003} 
Umeda, H., \& Nomoto, K., \nat {\bf 422} (2003) 871

\bibitem{umeda2005} 
Umeda, H. \& Nomoto, K., \apj {\bf 619} (2005) 427

\bibitem{wanajo2006} 
Wanajo, S., \apj (2006) in press (astro-ph/0602488)

\bibitem{wang2003} 
Wang, L., Baade, D., H\"oflich, P., \& Wheeler, J.C., \apj {\bf 592}
(2003) 457

\bibitem{wasserburg2000} 
Wasserburg, G.J., \& Qian, Y.-Z., \apj {\bf 529} (2000) L21

\bibitem{weiss2000} 
Weiss, A., Abel, T., \& Hill, V. (eds.), {\em The First Stars}
(Berlin: Springer) (2000)

\bibitem{weiss2004} 
Weiss, A., Schlattl, H., Salaris, M., et al., \aa {bf 422} (2004) 217

\bibitem{woosley1993} 
Woosley, S.E., \apj {\bf 405} (1993) 273

\bibitem{woosley2006} 
Woosley, S.E., \& Bloom, J., \annrev {\bf 44} (2006) in press

\bibitem{woo95} 
Woosley, S.E., \& Weaver, T.A., \apjs {\bf 101} (1995) 181

\bibitem{yoshii1981} 
Yoshii, Y., \aa {\bf 97} (1981) 280

\bibitem{zampieri2003} 
Zampieri, L., et al., \mnras {\bf 338} (2003) 711

\end{thebibliography}
\end{document}